\documentclass[11pt,a4paper]{article}
\usepackage{jcappub}
\pdfoutput=1

\usepackage{ruths_commands}
\usepackage{martins_commands}

\begin{document}

\title{Gravitational Waves sourced by\\ Gauge Fields during Inflation}

\author[a]{Martin Teuscher,}
\author[b]{Ruth Durrer,}
\author[a]{Killian Martineau,}
\author[a]{Aurélien Barrau}

\affiliation[a]{Laboratoire de Physique Subatomique et de Cosmologie, \\ 53 avenue des Martyrs, 38000 Grenoble, France}
\affiliation[b]{Département de Physique Théorique, Université de Genève,\\24 quai Ernest Ansermet, 1211 Genève 4, Switzerland}

\emailAdd{teuscher@lpsc.in2p3.fr}
\emailAdd{ruth.durrer@unige.ch}
\emailAdd{martineau@lpsc.in2p3.fr}
\emailAdd{barrau@in2p3.fr}

\abstract{We study the inflationary gravitational wave background induced by Abelian gauge fields generated by non-minimal kinetic and axial couplings to the inflaton. We show that, up to slow-roll corrections, for coupling functions that share the same dependence on conformal time, the gravitational wave spectrum is nearly scale invariant. We also derive its amplitude for generic gauge field coupling parameters, within the slow-roll approximation. The coupling values and the scale of inflation for which the induced gravitational wave background is observable, while ensuring that back-reaction on the inflationary dynamics remains negligible, are calculated. We find that a sizeable axial coupling can boost this secondary gravitational wave signal above the standard inflationary background. In the course of our analysis, we also show how to analytically match tensor perturbations across an arbitrary number of eras with different equations of state.}


\maketitle

\section{Introduction}
\label{s:intro}

The present Universe is permeated by magnetic fields on all scales, ranging from stars~\cite{Donati09}, to galaxies~\cite{Wielebinski05,Beck:2013it}, clusters of galaxies~\cite{Bohringer:2016aoa,Osinga:2022tos,Pignataro:2025ntd},  and filaments~\cite{Vernstrom:2021hru,Carretti:2024bcf}.
There is also indirect evidence that magnetic fields are present even in galactic voids~\cite{Neronov:2010b}. 
Especially, this last observation, along with the presence of magnetic fields in relatively high redshift galaxies ~\cite{Bernet08}, lead to the reasonable hypothesis that magnetic fields have a primordial origin (see, however, Ref.~\cite{Garg:2025mcc} for an alternative possibility). 

Primordial magnetic fields can be generated during phase transitions in the early universe due to the turbulent motion in the charged cosmic plasma. Such magnetic fields have a very short coherence length and a very blue power spectrum limited by causality~\cite{Durrer:2003ja}. Magnetic fields of primordial origin can also be generated during inflation, in which case they may have larger coherence scales and less blue spectra~\cite{Martin:2007ue,Durrer:2013pga,Subramanian:2015lua,Vachaspati:2020blt,Durrer:2022emo}. 
In particular, helical fields are interesting as their coherence scale can grow significantly after their generation due to the inverse cascade phenomenon~\cite{Campanelli:2007tc,Durrer:2013pga,Caprini:2014mja,Fujita:2019pmi}.

For gauge fields to be generated out of the vacuum in an expanding universe, they have to be non-minimally coupled either to curvature or to the inflaton. During slow-roll inflation these couplings are all equivalent (as we briefly show in \appx\ref{a:phiR}). Nevertheless, direct coupling of the inflaton to the $F^2$ term (so called kinetic coupling) is constrained in order to prevent a phase of strong coupling early during inflation~\cite{Demozzi:2009fu}. Interestingly, axial couplings are not affected by this consistency issue. Ways to avoid the strong coupling problem have been proposed in~\cite{Caprini:2014mja,Fujita:2019pmi}. In most cases it is found that the consistency constraints coming from strong coupling, joint with the ones ensuring the absence of back-reaction, only allow for rather blue spectra of magnetic fields after inflation~\cite{Durrer:2013pga,Subramanian:2015lua,Caprini:2014mja,Durrer:2022emo,Durrer:2023rhc}.

Electric fields  which decay exponentially fast in the charged plasma after inflation and small scale magnetic fields are rapidly damped by diffusion~\cite{Durrer:2013pga,Subramanian:2015lua}. However, even in the absence of helicity, an inverse cascade phenomenon is observed in magnetohydrodynamics (MHD) simulations~\cite{Brandenburg:2014mwa} and is reasonably well understood~\cite{Durrer:2013pga,Subramanian:2015lua}. This  leads to a less 
pronounced damping than expected from simple hydrodynamical arguments. With this, several inflationary magnetic field production mechanisms actually lead to fields which are promising candidates for the large scale cosmological magnetic fields present in voids and filaments.

In this paper we study the following problem. The energy momentum tensor generated by the electromagnetic field during inflation has an anisotropic stress with a transverse-traceless component. This component generates gravitational waves (GWs) that sum up with the usual gravitational waves generated by the amplification of vacuum fluctuations of the metric. We want to determine and characterize these secondary gravitational waves. During inflation these gravitational waves are super-horizon and are not oscillating, hence they should rather be called metric tensor perturbations but we shall use the term ``gravitational waves" since after inflation they will enter the horizon and lead to a gravitational wave background that we investigate in this study.
We neglect the possibility of additional GW generation during reheating that is very model-dependent and has recently been studied in Ref.~\cite{Maiti:2025cbi}. Also gravitational waves generated during the radiation era sourced by inflationary magnetic fields have been studied in the past, see~\cite{Atkins:2025pvg} for a recent paper. These additional GW background will therefore not be the topic of our work.

Contrary to previous work, where the induced GW background has been studied for purely axial~\cite{Barnaby:2011vw} or dominantly axial~\cite{Caprini:2014mja} coupling, here we study the full allowed range of kinetic and axial couplings within the slow-roll approximation. Contrary to a recent numerical study of a specific model of axion inflation~\cite{vonEckardstein:2025oic}, we find that there is a considerable inflationary parameter space that leads to a detectable GW background from gauge fields and that is safe from back-reaction.

In Section \ref{s:electromag} we discuss the generation of $U(1)$ gauge fields during inflation. These results are not new but we
present them for arbitrary slow-roll couplings and we study the limits imposed by back-reaction. We then derive analytical formulae for the induced anisotropic stress spectrum in full generality. In previous work, only the electric field~\cite{Caprini:2014mja} or only either kinetic~\cite{Martin:2007ue,Subramanian:2009fu} or axial~\cite{Durrer:2010mq,Barnaby:2011vw} couplings have been considered. In \sect\ref{s:GWprod} we compute the production of secondary gravitational waves due to these gauge fields. We also study the subsequent evolution of the gravitational waves through the radiation and matter dominated eras. To this aim, we develop a general formalism to transit from one era of constant $P/\rho =w$ to the next, that is applicable to an arbitrary number of eras. In \sect\ref{s:GWres}, we present and discuss the resulting power spectrum and energy density and in \sect\ref{s:con} we conclude. Several technical computations as well as  details about our notations are deferred to appendices.

\bigskip
\bigskip
\bigskip
\bigskip

\alinea{Notation}\\
\noindent
We consider a spatially flat background with metric
\be
\label{eq:FRLW-metric}
\dd s^2 =a^2(-\dd\tau ^2+\de_{ij}\dd x^i \dd x^j) \mcomma
\ee
where $a$ is the scale factor and $\ta$ is conformal time, related to physical time $t$ through $\dd t = a \dd \ta$. A prime (resp. overdot) denotes a derivative with respect to $\ta$ (resp. $t$). The physical Hubble parameter is $H=\flaf{\dot a}{a}$, while the conformal Hubble parameter is $\HH = \flaf{a'}{a} = aH$. As $x^i$ refer to comoving coordinates, $k$ refers to a comoving wavenumber, related to the physical wavenumber $k\subsc{phys}$ through $k = a k\subsc{phys} = a_0 k\subsc{phys,0}$ , where the label 0 indicates evaluation at the present time. Whenever slow-roll is invoked, we take the origin of (conformal) time such that $\ta <0$ during inflation and $a \simeq -\flaf{1}{(H\ta)}$ with $H\simeq \text{constant}$.
We define the reduced Planck mass by $\planckmass =1/\sqrt{8\pi G}$. Bold face letters denote spatial 3D vectors (although we sometimes omit it). We work in  Coulomb gauge, for which the quantized gauge field potential  $A_\mu$ is such that $A_0=\dr_jA^j=0$ and can hence be expanded as
\be
    \label{eq:quantum-expansion}
    \bm{A}(\xb,\tau) = \kint[\kb] \frac{1}{\sqrt{2|\kb|}} \sum_{\lambda=\pm}\left[\bm{\eps}^\lambda_{\kb} A^\lambda_{\kb}(\ta) \hat{a}^\lambda_{\kb} e^{+i\kb\cdot\xb} + \text{h.c.}\right]  \mcomma
\ee
with $[\hat{a}^\lambda_{\kb},(\hat{a}^{\lambda'}_{\kb'})^\dagger] = (2\pi)^3\delta^{\lambda\lambda'}\delta^{(3)}(\kb-\kb')$. Conventions for the helicity modes $\eps_k^\pm$ of the gauge field, as well as polarization tensors for GWs, are detailed in \appx\ref{a:conv}.

\section{Gauge field generation during inflation}
\label{s:electromag}
In four space-time dimensions, gauge fields are conformally coupled and are therefore -- contrary to scalar and tensor fluctuations -- not generated by the simple expansion of the Universe. For gauge fields to be excited during inflation, they have to be non-minimally coupled either to the curvature or to the inflaton. During slow-roll inflation, both options are in fact equivalent (c.f. \appx\ref{a:phiR} for details). In this paper we couple the gauge field $A_\mu$ to the inflaton $\phi$ and consider a simple $U(1)$ gauge field which may later become the electromagnetic field. We therefore work with the action
\be
\label{eq:action}
S  = \int \dd[4] x \sqrt{-g}\left[\pref{2}\planckmass^2 R -\frac{1}{2}\dr_\mu\phi\dr^\mu\phi -V(\phi) - \frac{1}{4}(1+i_1(\phi))F_{\mu\nu}F^{\mu\nu} - \frac{1}{4}i_2(\phi)F_{\mu\nu}\tilde F^{\mu\nu} \right] \mcomma
\ee
where $R$ is the Ricci scalar, $F_{\mu\nu}=\nabla_\mu A_\nu - \nabla_\nu A_\mu$ is the electromagnetic field tensor and $\tilde F^{\mu\nu} = \pref{2\sqrt{-g}}\epsilon^{\mu\nu\al\bt} F_{\al\bt}$ is its dual, $\epsilon^{\mu\nu\al\bt}$ is the Levi-Civita symbol in four dimensions. It has been shown in the literature that these two non-minimal couplings are actually of the most general form, providing the theory satisfies realistic assumptions \cite{Fleury:2014qfa}. The {\it kinetic} coupling $1+i_1$ modifies the canonical kinetic term of the gauge field and acts as a renormalization of the electric charge, $e\subsc{ren} = \flaf{e}{\sqrt{1+i_1}}$. In order to prevent strong coupling to charged particles, we request that $1+i_1$ never becomes very small~\cite{Demozzi:2009fu}. The {\it axial} coupling $i_2$ allows for the possibility of generating {\it helical} fields, as it acts with opposite signs on both polarizations. The generation of helical fields is motivated by the possibility of an inverse cascade process in the later radiation era, which can significantly increase their coherence length to cosmological scales \cite{Campanelli:2007tc,Brandenburg:2017rcb}.

\subsection{Sub- and super-horizon solutions of the equation of motion}
\label{sub:EM-eom}

Throughout this paper, the generation of electromagnetic fields (for convenience, we will use the same denominations as for the actual $U(1)\subsc{em}$ field) is assumed to remain small enough to be treated perturbatively. In particular, it is supposed to trigger no significant back-reaction on the evolution of the inflaton field. Several papers have  studied back-reaction, see Refs. \cite{Watanabe:2009ct,Domcke:2020zez,Durrer:2023rhc,Figueroa:2024rkr} for a non-exhaustive list. While some have found that the inflaton evolution is somewhat modified, leading to a prolonged inflationary phase~\cite{Durrer:2023rhc,Figueroa:2024rkr} due to the additional damping of the inflaton kinetic energy by its coupling to the gauge field, the energy momentum tensor of the gauge field remains typically very subdominant as exponential couplings are usually required for this back-reaction to become significant~\cite{Watanabe:2009ct}. 

In Ref.~\cite{Durrer:2010mq}, it has been shown that a purely axial coupling generically leads to a blue spectrum of magnetic fields with spectral index\footnote{Here $n_B$ is defined such that the power spectrum scales as $P_B(k)\propto k^{n_B-3}$ and the total power per log interval scales as $\PP_B(k)\propto k^3P_B(k)\propto k^{n_B}$.} $n_B=4$, and that back-reaction is negligible if the coupling is not too large. As we shall see, including also kinetic coupling allows for different spectral indices.

In our treatment we leave the inflaton potential $V(\phi)$ unspecified -- apart from requiring that it generates a slow-roll phase lasting sufficiently long. In  Coulomb gauge, $A_0 = \partial_j A^j = 0$, and the equation of motion for the field $A_\mu = (0,\bm{A})$ reads
\be
\bm{A}'' -\nabla^2 \bm{A} + \frac{i_1'}{1+i_1}\bm{A}' - \frac{i_2'}{1+i_1} \nabla \cross \bm{A}=0 \mcomma
\ee
where $i_n' \equiv \phi' \partial_\phi i_n$ and $\nabla^2 = \de^{ij}\partial_i \partial_j$ denotes the flat space Laplacian. The quantum expansion \eqref{eq:quantum-expansion} for $A_\mu$ yields the following equation of motion for the polarized mode functions $A_k^\lambda$ in Fourier space: 
\begin{equation}
\label{eq:eom-for-A}
    A^{\pm \prime\prime}_k + \frac{i_1'}{1+i_1} A^{\pm\prime}_k +\left( k^2 \mp k\frac{i_2'}{1+i_1} \right) A^\pm_k =0 \mperiod
\end{equation}
In this equation of motion it is manifest that $i_2$ affects the two polarizations with opposite signs, while the $i_1$ coupling is polarization independent. 

As we have mentioned earlier, the kinetic term of $A_\mu$ is not canonically normalized. This motivates the definition of a canonical auxiliary field
\be
\cA\symbs(\ta) = \sqrt{1+i_1(\ta)} A\symbs(\ta) \mcomma
\ee
for which \Eq\eqref{eq:eom-for-A} takes the simpler form
\be
\label{eq:eom-nice-form}
    \cA^{\pm \prime\prime}_k  +\left( k^2 \mp k\frac{i_2'}{1+i_1} - \frac{(\sqrt{1+i_1})''}{\sqrt{1+i_1}} \right) \cA^\pm_k =0 \mperiod
\ee

The reader familiar with standard cosmological perturbation theory may recognize the similarity between \Eq\eqref{eq:eom-nice-form} and the Mukhanov-Sasaki equation $u_k''+\left(k^2-\frac{z''}{z}\right)u_k = 0$. We refer to \appx\ref{a:mukha} for a more comprehensive discussion on the common grounds and differences between these two equations. 

\smallskip
We separately explore the sub-horizon $k\gg \bigO(\HH)$ and super-horizon $k\ll \bigO(\HH)$ behaviors of the solutions to \Eq\eqref{eq:eom-nice-form}. For that, we can either solve the differential equation, then asymptotically expand its solution, or simplify the equation using these asymptotics then solve the reduced equation. However, we want to draw attention to the fact that these two operations (solving and taking the limit) do not commute in general, and there is no guarantee that both schemes will lead to the same solution on the whole interval of definition. This will be illustrated below. Therefore, we choose the path of keeping the equation of motion exact and  we  perform the expansion only after obtaining the full solution.

However, \Eq\eqref{eq:eom-nice-form} does not have analytic solutions for arbitrary coupling functions $i_1$ and $i_2$. Motivated by pursuing an analytic resolution further without relying on numerical modeling of one specific coupling function, we restrict ourselves to slow-roll inflation. Taking $\tau<0$, we consider that $\phi(\ta)$ varies slowly enough for the following assumptions on $i_1(\phi)$ and $i_2(\phi)$ to hold
\begin{equation}\label{e:i1i2power1}
\dv{\ln(1+i_1)}{\ln(-\ta)}\simeq \const \equiv \gaun \spliteq 
\frac{-\tau\, i_2'}{(1+i_1)} \simeq \const \equiv \gade \mcomma
\end{equation}
or, equivalently,
\begin{eqnarray}
    \label{eq:drondphi-i}
    \partial_\phi i_m = \gamma_m(1+i_1) \frac{H}{\dot{\phi}}\quad \text{for } m\in\{1, 2\}\mperiod
\end{eqnarray}
During slow-roll $\phi$ can be considered to vary monotonically with time. Therefore the functions $i_m(\phi(\tau))$ can also be interpreted as functions of time. A typical realization of the chiral coupling is standard axion inflation where $i_1= \ga_1=0$ and $i_2 = (\al/f)\phi$ where $f$ is the axion decay constant and $\al$ is a dimensionless parameter. In this case
\begin{equation}
\ga_2 = \partial_\phi i_2 \frac{\dot\phi}{H} = \frac{\al}{f}\frac{\dot\phi}{H}= 2\xi \mcomma
\end{equation}
see e.g~\cite{Barnaby:2011vw}, Eqs. (2.1) and (2.10). In this case, one can also include a non-trivial kinetic coupling $i_1$ but it has to be slowly rolling, i.e. $\left|\ga_1\right| \ll 1$. A non-trivial kinetic coupling is usually encountered in models searching for variations of the fine structure constant with time, as a varying kinetic term introduces a modification of the $\mathrm{U}(1)$ electric charge \cite{Demozzi:2009fu}.

The equations \eqref{e:i1i2power1} are solved by\footnote{Some authors (see e.g. \cite{Martin:2007ue}) rather define an index $n$ such that $1+i_1 \propto a^{2n}$, then relate $a$ to $\ta$ using the slow-roll parameter ($a\propto (-\ta)^{-1}$ at lowest order, $a\propto (-\ta)^{-1-\ep}$ at the next-to-leading order). This modifies the definition of $\gaun$ to $\gaun=-2n(1+\ep)$. With our definition of $\gaun$ as the variation of $i_1$ with $\ta$, the slow-roll parameter does not introduce a correction to $\gaun$ and the coefficients in the equation of motion \eqref{eq:eom-with-g1-g2} remain insensitive to the value of $\ep$.} 
\begin{equation} \label{e:i1i2power}
1+i_1 \simeq \left(\frac{\tau}{\tau\ssend}\right)^\gaun \spliteq  i_2 \simeq - \frac{\gade}{\gaun}\left[\left(\frac{\tau}{\tau\ssend}\right)^\gaun - 1\right]\mperiod
\end{equation}
Here, $\ta\ssend$ denotes the time at the end of slow-roll (and of inflation, as we consider a simplified scenario with instantaneous reheating), and we have chosen the constants of integration such that $1+i_1(\ta\ssend) = 1$ and $i_2(\ta\ssend)=0$, so that standard electromagnetism is recovered after inflation.\footnote{In fact, one may fix $i_2$ to any constant value after the end of inflation. When $i_2$ is constant, the term $F\munu\tilde F\upmunu$ in the Lagrangian becomes a surface term that has no influence on the bulk equations of motion.} This parametrization has been studied in e.g. \cite{Caprini:2014mja,vonEckardstein:2025oic,Sobol:2020lec,Dimopoulos:2024jnv} for $i_1$ ($\gaun = -2n$, $\gaun = -2f_1$), and in e.g. \cite{Durrer:2010mq,Barnaby:2011vw,Caprini:2014mja} for $i_2$ ($\gade = 2 \xi$).
As an illustration of the validity of $\gaun,\gade\simeq \const$, the case developed in \cite{Barnaby:2011vw} of $i_2(\phi) \propto \phi$ yields $\gade \propto \flaf{\dot{\phi}}{\sqrt{V(\phi)}}\propto \sqrt{\ep}$ with $\ep$ the first slow-roll parameter, which is constant at first order in  slow-roll.

It might seem that the above restriction on the functions $i_1$ and $i_2$ is rather severe. But in Ref.~\cite{Giovannini:2021thf} it is shown that it has very little influence on the generated gauge field spectra.

Finally, while the sign and magnitude of $\gade$ are not restricted {\it a priori}, we must limit our analysis to the case $\gaun \geqslant 0$. The case $\gaun <0$ leads to $1+i_1 \to 0$ during the early phase of inflation, inducing a regime of strong coupling with very large electric charge. In this strong coupling regime we cannot trust our analysis~\cite{Demozzi:2009fu}. 

\bigskip
Under these assumptions, \Eq\eqref{eq:eom-nice-form} becomes
\begin{equation}
    \label{eq:eom-with-g1-g2}
     \cA^{\pm \prime\prime}_k  +\left( k^2 \pm k \frac{\gade}{\ta} + \frac{\gaun(2-\gaun)}{4\tau^2} \right) \cA^\pm_k =0 \mperiod
\end{equation}

This equation naturally introduces a notion of ``electromagnetic (comoving) horizon". This horizon refers to the mode $k$ at a given time $\ta$ for which the $k^2$ term in the brackets of \Eq\eqref{eq:eom-with-g1-g2} becomes subdominant, and thus at which we expect the behavior of the solution to change due to the couplings. More explicitly, this mode is
\begin{equation}
    \label{eq:def-electro-horizon}
    \widetilde{k}_h(\ta)  \equiv \pref{-\ta}\left(\frac{|\gade|}{2} + \pref{2}\sqrt{\gade^2 + \left|\gaun(2-\gaun)\right|}\right) 
\end{equation}
(note the introduction of absolute values that keep $\widetilde{k}_h$ positive). In the remainder of this study however we will use approximations which hold only for $-k\ta \ll 1$, so we also introduce
\begin{equation}
    \label{eq:def-gamma3-with-min}
   k_h(\ta) \equiv \pref{-\ta}\min\left(1\, ; \frac{|\gade|}{2} + \pref{2}\sqrt{\gade^2 + \left|\gaun(2-\gaun)\right|}\right) \equiv -\frac{\gamma_3}{\ta} = \ga_3\HH\mperiod
\end{equation}
Although the true electromagnetic horizon  is given by \eqref{eq:def-electro-horizon}, we do not expect the approximations derived below to be accurate for $\HH < k < \widetilde{k}_h$ if $\widetilde{k}_h \gg \HH$ and hence we will only use the latter definition \eqref{eq:def-gamma3-with-min} for practical applications. We similarly define the ``electromagnetic horizon" crossing time $\ta_h(k)$ by 
\begin{equation}
    \label{eq:def-electro-horizon-crossing-time}
    \ta_h(k) \equiv -\frac{\gatr}{k}\qquad \text{i.e.}\qquad k_h(\ta_h(k))=k\mperiod
\end{equation}

The generic solution of the equation of motion \eqref{eq:eom-with-g1-g2} with constant $\ga_1$ and $\ga_2$ is
\begin{equation}
     \cA\symbs(\ta) = \lambda\symbs\whit_{\mp i \frac{\gade}{2}, \frac{1-\gaun}{2}}(2 i k \ta) +\mu\symbs\whitM_{\mp i \frac{\gade}{2},\frac{1-\gaun}{2}}(2 i k \ta)\mcomma
\end{equation}
where $\whit_{\kappa, \mu}(z)$ and  $\whitM_{\kappa,\mu}(z)$ are the Whittaker functions~\cite{Abramo, nist-whittaker}.
To determine the coefficients $\lambda\symbs$ and $\mu\symbs$, we impose vacuum Bunch-Davies initial conditions with positive frequency for the gauge field at early times where the mode $k$ is deeply sub-horizon, i.e. we want $\cA\symbs(\ta)$ to behave as
\begin{equation}
\label{eq:bunch-davis-solution}
   \frac{1}{\sqrt{2k}} e^{-ik \ta}\, \quad \text{for}\quad \left|k\tau\right|\gg 1 \mperiod
\end{equation}
 In this limit, $\whitM_{\kappa,\mu}(z)$ contains terms with both positive and negative frequencies \cite{Abramo}, and must thus be discarded. However, and we emphasize this somewhat overlooked point, the limit of the remaining solution reads
\begin{equation}\label{e:inital}
    \frac{1}{\sqrt{2k}}\cA\symbs \underset{k|\ta|\gg 1}{\sim} \frac{\lambda\symbs}{\sqrt{2k}}e^{-ik\ta} (2i k \ta)^{\mp i \gade/2} =
\frac{\lambda\symbs}{\sqrt{2k}}e^{-ik\ta \mp (\pi\gade/4)\mp i(\gade/2)\ln(|2 k \ta| )}
\end{equation}
(where the complex logarithm is defined on $\mathbb{C}\backslash\mathbb{R}_-$). It thus appears that this solution {\it cannot} be formally matched to the vacuum expression \eqref{eq:bunch-davis-solution}. This is a direct consequence of the fact that solving the differential equation \Eq\eqref{eq:eom-with-g1-g2} and taking the limit $k\ta \to -\infty$ are two non-commutative operations. Indeed, if one would have removed the last two terms in the brackets of \Eq\eqref{eq:eom-with-g1-g2}, the positive frequency solution would precisely be the vacuum solution \eqref{eq:bunch-davis-solution}.

Nevertheless, we can bypass this formal difficulty by imposing the vacuum solution to $|\cA\symbs|^2$ rather than $\cA\symbs$ directly (but after discarding the negative frequency solution), namely
\begin{equation}
    \left|\frac{1}{\sqrt{2k}}\cA\symbs\right|^2 \underset{k|\ta|\gg 1} {\sim} \frac{1}{2k} \mperiod
\end{equation}
This leads to $|\lambda\symbs|^2 = \exp(\pm \pi \gade/2)$, from which we set\footnote{As these are stochastic fields, we will be only interested in real, quadratic averages. Hence, picking any $\lambda\symbs = e^{i\th}e^{\pm \pi \gade/4}$ with $\th\in\mathbb{R}$ describes the same physics, so we conveniently set $\th=0$.} $\lambda\symbs = \exp(\pm \pi \gade/4)$, and therefore 
\begin{equation}
\label{eq:final-whittaker-solution}
    \cA\symbs(\ta) = e^{\pm \pi \gade/4}\whit_{\mp i \frac{\gade}{2}, \frac{1-\gaun}{2}}(2 i k \ta)\mperiod
\end{equation}
In the super-horizon limit, $-k\ta\to 0^+$, this expression becomes for $\gaun\notin \mathbb{N}^*$ 
\begin{equation}
\label{eq:super-horizon-whittaker}
     \cA\symbs = e^{\pm \pi \gade/4}\left[\frac{\Ga(1-\gaun)}{\Ga\left(1-\frac{\gaun}{2}\pm i \frac{\gade}{2}\right)}(2i k \tau)^{\gaun/2} + \frac{\Ga(\gaun-1)}{\Ga\left(\frac{\gaun}{2}\pm i \frac{\gade}{2}\right)}(2i k \tau)^{1-\gaun/2}  + \bigO(\text{subdominant})\right] \mperiod
\end{equation}
For $\gaun\in\nat$ a series expansion still exists, for example at $\gaun=1$ we obtain in the super-horizon limit
\begin{equation}
\label{eq:super-horizon-whittakerg11}
     \cA\symbs = - e^{\pm \pi \gade/4}(2ik\tau)^{1/2}\left(\frac{\ln(2ik\ta) + 2\ga_E+\psi(\frac{1}{2}\pm i\frac{\ga_2}{2})}{\Ga(\frac{1}{2}\pm i\frac{\ga_2}{2})}\right)  + \bigO(|k\ta|^{3/2}|\ln(-k\ta)|)\mperiod
\end{equation}
Here $\ga_E$ is Euler-Mascheroni's constant, $\ga_E\simeq 0.577216$, and $\psi(z) = \flaf{\Ga'(z)}{\Ga(z)}$ is the digamma function.
Although the expansion \eqref{eq:super-horizon-whittaker} agrees with \Eq (2.6) in~\cite{Caprini:2014mja} in the limit $\ga_2\gg 1$ (using some identities for the $\Ga$-function), it exhibits  spurious divergences at integer values of $\gaun$, that the true solution does not have. In order for the approximation to remain faithful, a careful treatment involving the subdominant terms is in fact required. We postpone this analysis to Section \ref{sub:Em-spectra}, as the treatment of the magnetic and electric fields differs somewhat.

Lastly, to relate our analysis to previous studies, let us provide $\cA\symbs $ in simpler scenarios. 
When $\gade = 0$, the Whittaker functions simplify to Hankel functions and \Eq\eqref{eq:final-whittaker-solution} reads
\begin{equation}
    \cA\symbs(\tau) = \zeta\sqrt{\frac{\pi}{2}}\sqrt{-k\ta}H_{\frac{1-\gaun}{2}}^{(1)*}(-k\ta) \underset{k|\ta|\ll 1}{\sim}\zeta' \frac{1}{\sqrt{\pi}}\Ga\left(\frac{1-\gaun}{2}\right)\left(-\frac{k\ta}{2}\right)^{(1-|1-\gaun|)/2}\mcomma
\end{equation}
where $\zeta,\zeta'\in U(1)$ are irrelevant phase factors and $H_\nu$ is the Hankel function of order $\nu$, see~\cite{Abramo}. If $\ga_1$ is an even integer, the Whittaker functions simplify to Coulomb wave functions of  order $\ga_1/2-1\in\nat$, and we find that \Eq\eqref{eq:final-whittaker-solution} agrees with the solution in~\cite{Caprini:2014mja}. In the previous literature, $\ga_1=0$ has been studied in~\cite{Durrer:2010mq,Sorbo:2011rz,Barnaby:2011vw} where in \cite{Sorbo:2011rz,Barnaby:2011vw} the limit $\ga_2\gg 1$ was considered. In~\cite{Caprini:2014mja}, both $\ga_1\neq 0$ and $\ga_2\neq 0$ have been studied but again in the limit $\ga_2\gg 1$. Here we consider arbitrary values of $\ga_1$ and $\ga_2$ with the only restriction that $0\leqslant\ga_1< 4$. The lower limit is to avoid strong coupling and the upper one to avoid infrared divergences, as we shall see below.

\subsection{(Anti)symmetric electromagnetic spectrum}
\label{sub:Em-spectra}

Let us determine the symmetric and anti-symmetric power spectra of the electric and magnetic fields generated by these  couplings. They are defined in terms of the mode functions in the polarization basis of the quantum (or stochastic) field $\hat{X}(\kb,\ta)$ as 
\begin{equation}
\label{eq:inbody-correlation-from-mode-functions}
    S_X = \frac{1}{2k}\sum_{\lambda=\pm s}X_k^\lambda(\ta)X_k^{\lambda *}(\ta') \spliteq A_X = \frac{1}{2k}\sum_{\lambda=\pm s}(\lambda/s)X_k^\lambda(\ta)X_k^{\lambda *}(\ta')\mperiod
\end{equation}
Here $X_k^\la$ is the mode function of helicity $\la$ of the field $\hat{X}(\kb,\ta)$, and $s$ stands for its spin: $s=1$ for gauge fields, but we will use the same definitions with $s=2$ for the GW spectrum in Section \ref{s:GWprod}. We refer the reader to \appx\ref{a:conv} for a comprehensive definition.

The standard dimensionless\footnote{\label{foot:dimensionless} Here the word `dimensionless' is somewhat an abuse of speech. It actually means that $\PP_X$ and $\PP_X^A$ have the same dimension as $X^2$ in real space. } symmetric and anti-symmetric power spectra, $\PP_X$ and $\PP_X^A$, are then given by
\begin{equation}
\label{eq:powerspectrum-from-correlation}
    \PP_X(k,\ta) = \frac{k^3}{2\pi^2}S_X(k,\ta,\ta)\spliteq  \PP_X^A(k,\ta) = \frac{k^3}{2\pi^2}A_X(k,\ta,\ta)\mperiod
\end{equation}

We define the magnetic and electric fields associated with the gauge field $A_\mu$ by
\begin{equation}
    B_j(\bm{k}) = \pref{a} \epsilon^{jlm}\kb\supsc{phys}_l A_m(\bm{k}) = \pref{a^2} \epsilon^{jlm}\kb_l A_m \spliteq[,] E_j = -\pref{a}\dv{A_j}{t} = -\pref{a^2} \dv{A_j}{\ta}\mcomma
\end{equation}
where we use the convention \cite{maggiore_vol2} of raising and lowering spatial indices of perturbative quantities with the Kronecker delta, i.e. $E_j = E^j, B_j=B^j$ so that $\rho = (E_j E^j+B_jB^j)/2 \propto a^{-4}$ gives the correct scaling of the energy density with the expansion.\footnote{Another possibility is to define $B_j = \epsilon^{jlm}\kb\supsc{phys}_l A_m$, $E_j = -(1/a)\dv*{A_j}{t}$, then move indices with $g\munu$. Then, $B_j\propto 1/a$, $B^j\propto 1/a^3$ and we still obtain $B_jB^j \propto 1/a^4$.}

However, $E$ and $B$ are not the fields that directly contribute to the production of gravitational waves. The stress-energy tensor entering the Einstein equation is $T\munu \equiv -(\flaf{2}{\sqrt{-g}})\times \pdv*{\mathcal{L}\subsc{em}}{g\upmunu}$, and thus is altered by the modification of the gauge field kinetic term in \Eq\eqref{eq:action}. As studying the generation of these waves is the main purpose of this work, we define the adequate source fields 
\begin{equation}
    \B_j = \sqrt{1+i_1} B_j = \pref{a^2} \epsilon^{jlm}\kb_l \cA_m ~,\quad \E_j = \sqrt{1+i_1} E_j = -\pref{a^2} (1+i_1)^{1/2}\dv{[(1+i_1)^{-1/2}\cA_j]}{\ta}
\end{equation}
so that
\begin{equation}
   \label{eq:def-of-curl-B-E}
  \B\symbs = \pm \pref{a^2}|\kb|\cA\symbs~,\quad \E\symbs = -\pref{a^2}(1+i_1)^{1/2}\dv{[(1+i_1)^{-1/2}\cA\symbs]}{\ta} \mperiod
\end{equation}

We can now provide explicit expressions for $\PB$ and $\PE$. As we are interested in the super-horizon limit of these fields, it is tempting to insert directly the lowest non-trivial order of \eqref{eq:super-horizon-whittaker} into \eqref{eq:def-of-curl-B-E}. However, this leads to an unphysical divergence of the fields at $\gaun=0$ and $\gaun=1$, while the original Whittaker solution is perfectly smooth at these values. The divergence at $\gaun=1$ was already observed in \cite{Martin:2007ue}.  We now explain how to construct reasonably faithful approximations to the actual solution.\footnote{We still want to use an approximation because keeping the exact Whittaker function would make all the subsequent computations analytically untractable.} We  discuss the magnetic and electric fields separately because the time derivative involved in the definition of $\E\symbs$ requires a slightly different treatment. 

\begin{enumerate}[label={\it (\roman*)}]
    \item {\it Magnetic field.} The first term in the brackets of \Eq\eqref{eq:super-horizon-whittaker}, which dominates if $0\leqslant \gaun < 1$, is finite when $\gaun\to 0$ but diverges when $\gaun\to 1$. We have assessed that when adding the second term, also divergent in this limit, the total sum becomes smooth and fairly close to the true solution. Unfortunately, adding this new term introduces a new divergence when $\gaun\to 0$. This divergence can in turn be canceled by the addition of the third term in the expansion (proportional to $(2ik\ta)^{1+\gaun/2}$), but this regenerates a divergence when $\gaun\to 1$, etc. As truncating the series to a finite order always leaves one of the divergences, we have to truncate at a different order depending on whether $\gaun$ is close to $0$ or $1$. 
    Moreover, for $1\leqslant \gaun < 2$ the situation is similar upon exchanging the role of the first and second term of \Eq\eqref{eq:super-horizon-whittaker}, because the equation of motion is symmetric under $\gaun \to 2-\gaun$. Finally, if $\gaun\geqslant 2$, the second term is both dominant and divergence-free, hence it suffices as an approximation.
    
    \item {\it Electric field.} The situation for the electric field is somewhat different because the first term of \Eq \eqref{eq:super-horizon-whittaker} gives no contribution to $\E\symbs$, see \Eq\eqref{eq:def-of-curl-B-E}. If $\gaun \geqslant 1$ the second term  dominates and is divergence-free, hence sufficient. However it diverges when $\gaun\to 0$, so in this limit it must be supplemented with the third term of the expansion \eqref{eq:super-horizon-whittaker}, namely
    \begin{equation}
    \label{eq:whittaker-3rd-term}
        \cA_k^{\pm,\text{3rd term}} = \mp e^{\pm\pi\gade/4} \frac{ i \gade}{2\gaun}\frac{\Ga(1-\gaun)}{\Ga\left(1-\frac{\gaun}{2} \pm i \frac{\gade}{2}\right)}(2i k \ta)^{1+\gaun/2} \mperiod
    \end{equation}
However, this last term generates a new divergence at $\gaun=1$, so once again we must use different approximations whether $\gaun$ is close or not to $0$.
\end{enumerate}

We therefore use the following approximations when $k\ta\to 0^-$:
\begin{align}
\label{eq:Bk-approximation}
    \pref{\sqrt{2k}}\B\symbs &\simeq \pm    \frac{k}{a^2} \frac{e^{\pm\pi\gade/4}}{\sqrt{2k}} (2ik\ta)^{1/2-|1-\gaun|/2}\delta^\pm_B(\gaun,\gade) \mcomma \\
    \label{eq:Ek-approximation}
    \pref{\sqrt{2k}}\E\symbs &\simeq \frac{2ik}{a^2} \frac{e^{\pm\pi\gade/4}}{\sqrt{2k}} (2ik\ta)^{-\gaun/2}\delta^\pm_E(\gaun,\gade) \mcomma
\end{align}
where $\delta^\pm_B,\delta^\pm_E$ are piecewise continuous functions of $\gaun$ as well as smooth functions of $\gade$, that remain finite for all values of $\gaun$. In fact, far from $\gaun=0$ one has $\delta^\pm_E = \Ga(\gaun)/\Ga\left(\frac{\gaun}{2}\pm i \frac{\gade}{2}\right)$, while far from $\gaun=1$, $\delta^\pm_B = \Ga(|1-\ga_1|)/\Ga\left(\frac{1}{2}+\frac{1}{2}|1-\gaun|\pm i \frac{\gade}{2}\right)$. Their complete expressions can be found in \appx\ref{asub:EM-comput}.

This fixes the issue of using divergent approximations. Inserting these approximations for the fields and using
\Eqs\eqref{eq:inbody-correlation-from-mode-functions} and \eqref{eq:powerspectrum-from-correlation}, we obtain the following spectra,
\begin{align}
\label{eq:S-EB}
S\subsc{em}(k,\ta,\ta') &\equiv S_\B(k,\ta,\ta')+S_\E(k,\ta,\ta')   \\
\label{eq:S-Bonly}
S_\B(k,\ta,\ta')&\simeq \frac{k^{2-|1-\gaun|}}{2a^2(\ta)a^2(\ta')}(\ta\ta')^{1/2-|1-\gaun|/2}\cosh(\frac{\pi\gade}{2})\De_B(\gaun,\gade) \\
S_\E(k,\ta,\ta') &\simeq \frac{k^{1-\gaun}}{2a^2(\ta)a^2(\ta')}(\ta\ta')^{-\gaun/2}\cosh(\frac{\pi\gade}{2})\De_E(\gaun,\gade) \\
\label{eq:PB}
    \PB(k,\ta) &\simeq \frac{1}{4\pi^2}\frac{k^4}{a^4}(-k\ta)^{1-|1-\gaun|}\cosh(\frac{\pi\gade}{2})\De_B(\gaun,\gade)  \\
\label{eq:PE}
    \PE(k,\ta) &\simeq \frac{1}{4\pi^2}\frac{k^4}{a^4}(-k\ta)^{-\gaun}\cosh(\frac{\pi\gade}{2})\De_E(\gaun,\gade) \\
\label{eq:A-EB}
 A_\B(k,\ta,\ta') &= S_\B(k,\ta,\ta')\tanh(\frac{\pi\gade}{2}) \\
 A_\E(k,\ta,\ta') &= S_\E(k,\ta,\ta')\tanh(\frac{\pi\gade}{2}) \mcomma
\end{align}
and similarly for $\PB^A$, $\PE^A$, and $A\subsc{em}=A_\B + A_\E$.

The definition of $\De_B$, $\De_E$  can again be found in \appx\ref{asub:EM-comput}. We recall that these expressions are good approximations only for $k|\ta|\ll 1$.

From the power spectra for the magnetic and the electric field, one can also infer the mean energy density of the gauge field
\be
\label{eq:EB-energy-density}
\dv{\rho}{\ln k} = \frac{1}{2}\left(\PB(k,\ta)+\PE(k,\ta)\right)\mperiod
\ee
Eqs.~\eqref{eq:PB} to \eqref{eq:PE} show that $\PE$ and $\PE^A$ are  blue for $0 \leqslant \ga_1 < 4$ and scale invariant for $\ga_1=4$, whereas $\PB$ and $\PB^A$ are blue in all cases for $0 \leqslant \ga_1 \leqslant 4$, in accordance with \cite{Maiti:2025cbi,Caprini:2014mja}. This means that if their energy density remains subdominant during inflation, after inflation when small scale magnetic fields as well as the entire electric field are damped away, the fields become very subdominant. However, for sufficiently large $\ga_1$ and $\ga_2$, the subsequent inverse cascade can still render them interesting for the problem of large scale cosmological magnetic fields (see Ref.~\cite{Caprini:2014mja}). As we shall see in Section~\ref{s:GWprod}, they can also generate interesting gravitational waves. Furthermore, \Eq\eqref{eq:A-EB} shows that the presence of the axial coupling enhances one polarization over the other, depending on the sign of $\gade$. The subsequent gravitational waves will then also be strongly polarized.

\begin{figure}
    \centering
\includegraphics[width=0.49\linewidth]{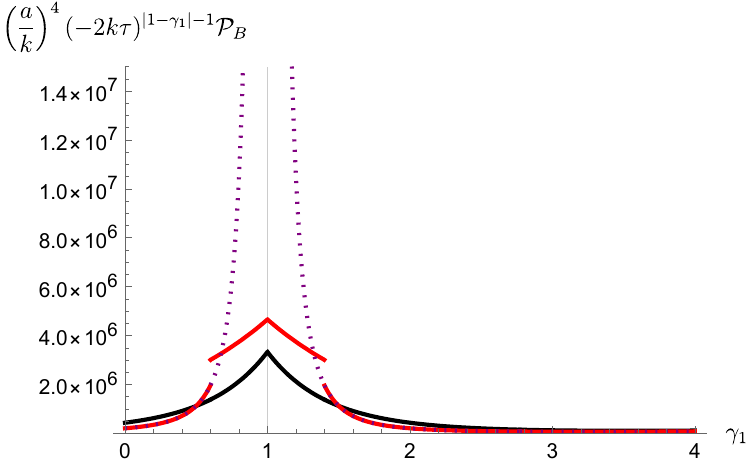}
\includegraphics[width=0.49\linewidth]{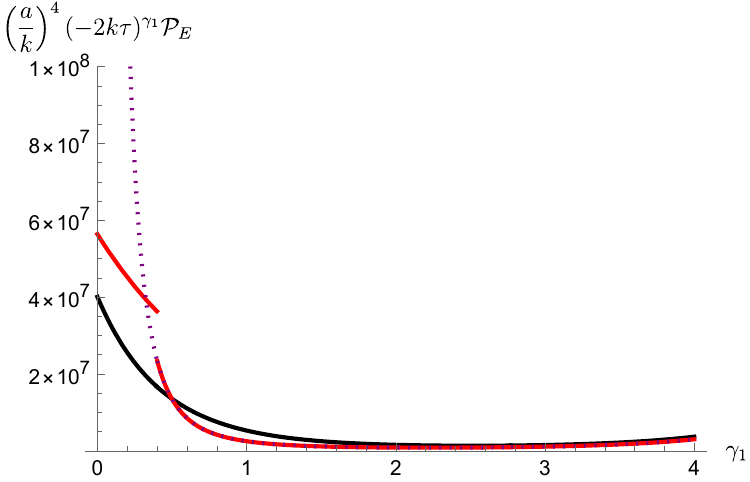}

    \caption{The scale independent pre-factors of the symmetric power spectra of magnetic and electric fields as a function of $\gaun$. Solid black lines correspond to the exact solution obtained from \eqref{eq:final-whittaker-solution};  solid red lines are the piecewise approximations \eqref{eq:PB} and \eqref{eq:PE}. Although these are not very accurate, they remain much closer to the original solution than the divergent approximation originating from the lowest order term of \eqref{eq:super-horizon-whittaker}, shown as dotted purple lines. The offset between the black and red lines at $\gaun=1$ (for $\PP_\B$) and $\gaun=0$ (for $\PP_\E)$ is due to having neglected the $\ln(2ik\ta)$ correction in \eqref{eq:Bk-approximation} and \eqref{eq:Ek-approximation}. Interestingly, the dependence on $\gaun$ is non-monotonic. Meanwhile, the dependence on $\gade$ is simpler and will be discussed in Appendix~\ref{asub:EM-comput}, see \fig\ref{fig:anisotropy-spectra}.  Here we have set $\gade=6, |k\ta|=10^{-2}$ and we choose $\delta=0.4$.} 
    \label{fig:electro-fields}
\end{figure}

 The magnetic field spectrum becomes scale invariant only for $\ga_1 = 6$. In this case, however, the electric field spectrum is red and requires an infrared cutoff, which is why we have restricted our analysis to $0\leqslant \gaun \leqslant 4$ in the first place. In order to test our approximations,  the  spectra are illustrated as a function of $\ga_1$ in \fig\ref{fig:electro-fields} for some values for $\ga_2$ and $k\tau$. The electric field dominates everywhere except for $\ga_1\sim 1$.  In this regime, while providing the right order of magnitude, our approximation is, however, rather poor.

\subsection{Back-reaction}
\label{sub:backreaction}

To test our slow-roll approximation, let us  discuss the significance of back-reaction of the gauge fields on the {inflationary dynamics}. One key assumption for the validity of our approach is that the background spacetime remains close to the quasi-De Sitter solution. This assumption may break down if the gauge fields trigger strong back-reaction {on the inflaton field or on the background metric}. This is quite a complex topic and we do not aim at accounting for this effect (see e.g. \cite{Durrer:2023rhc,Figueroa:2024rkr} for a more complete treatment), but  we derive consistency constraints to ensure that it is always subdominant, so that our analysis remains self-consistent.

Back-reaction can take place at two different levels. Firstly, it can modify the inflaton dynamics. In the presence of gauge fields, the inflaton equation of motion becomes
\begin{align}
    \ddot{\phi} + 3H\dot{\phi}+ \partial_\phi V &= -\frac{1}{4}\partial_\phi i_1 \avg{F\munu F\upmunu} -\frac{1}{4}\partial_\phi i_2 \avg{F\munu \tilde{F}\upmunu} \\
    &=\frac{1}{1+i_1}\left[\pref{2} \partial_\phi i_1 \avg{\bm{\E}^2-\bm{\B}^2} + \partial_\phi i_2 \avg{\bm{\B}\cdot\bm{\E}}\right] \mperiod
\end{align}
On the right hand side we have introduced expectation values, since the gauge fields are quantized whereas the inflaton field is classical. The gauge fields can potentially drive the inflaton out of slow-roll, in a regime where our analysis breaks down. We therefore impose the (sufficient) first condition
\begin{equation}
\label{eq:backreaction-1st-condition}
   \frac{1}{1+i_1}\left|\frac{1}{2} \partial_\phi i_1 \avg{\bm{\E}^2-\bm{\B}^2}\right| +  \frac{1}{1+i_1}\left|\partial_\phi i_2 \avg{\bm{\E}\cdot\bm{\B}}\right| \ll \left|3H\dot{\phi}\right| \simeq \left|\partial_\phi V\right| \mperiod
\end{equation}
Secondly, back-reaction on the expansion dynamics can occur if the energy density of the gauge fields becomes a sizeable fraction of the total energy density, as it would modify the Friedmann equation and again drive the system out of slow-roll. We thus impose the second condition
\begin{equation}
\label{eq:backreaction-2nd-condition}
    \rho \ll \rho_\phi \spliteq[i.e.] \frac{1}{2(1+i_1)}(\bm{\B}^2+\bm{\E}^2) \ll 3\planckmass^2 H^2 \mperiod
\end{equation}

\begin{figure}
    \centering
    \includegraphics[width=0.49\linewidth]{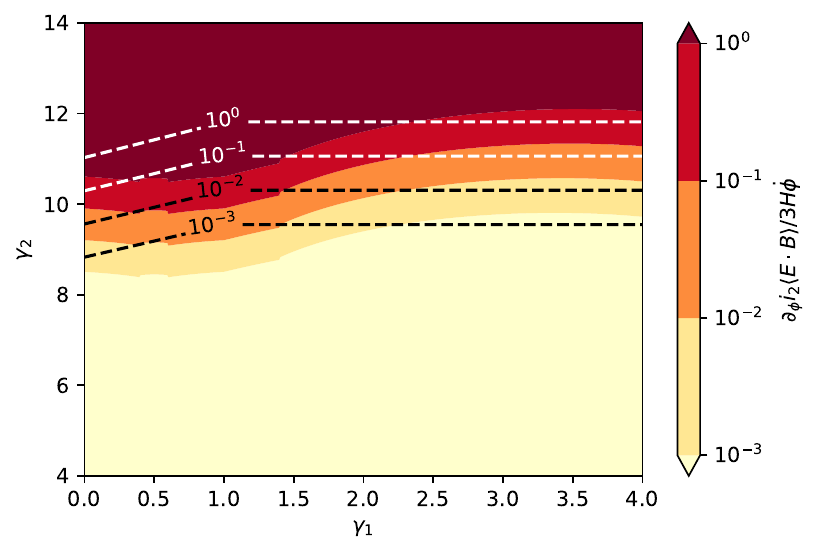}
    \includegraphics[width=0.49\linewidth]{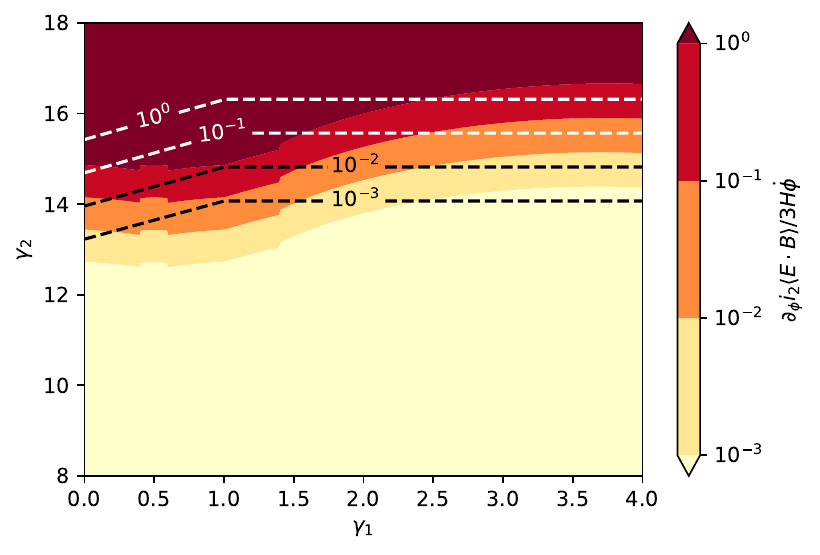}
    \caption{Constraints on $\gaun,\gade$ required by the self-consistency relation at the level of the equation of motion, which is the most stringent. Dashed lines show the constraints computed with the approximation~\eqref{eq:backreaction-large-g2} instead of the complete formula. We have considered a slow-roll parameter $\ep=0.1$; the inflation energy scale is set at $T\ssend = 10^{15}$GeV (left plot) and $T\ssend = 10^{12}$GeV (right plot). $H\ssend(T\ssend)=\flaf{\HH\ssend}{a\ssend}$ is given by \Eqs \eqref{eq:a-end}--\eqref{eq:hub-end} and is respectively $\simeq \SI{2.2e12}{\GeV}$  and $\simeq \SI{2.2e9}{\GeV}$.} 
    \label{fig:self-consistency-constraints}
\end{figure}

Using the expansion \eqref{eq:quantum-expansion}, we find
\begin{align}
\avg{\bm{\E}^2 \pm \bm{\B}^2} &= \int_0^{k_h} \frac{\dd k}{k}\left(\PP_\E(k)\pm \PP_\B(k)\right) \mcomma\\
    \avg{\bm{\B}\cdot\bm{\E}} &= \sum_{\lambda=\pm} \int_0^{k_h} \frac{\dd k}{k} \frac{k^3}{2\pi^2 }\pref{4k}\left(\B_k^\lambda\E_k^{\lambda *} + \E_k^\lambda\B_k^{\lambda *}\right) \mperiod
\end{align}
We have prematurely introduced the UV-cutoff $k_h = \gatr \HH$ according to the argument of Section \ref{subsub:aniso-cutoff} that sub-horizon modes do not contribute to the overall energy density. Inserting \Eqs\eqref{eq:Bk-approximation}, \eqref{eq:Ek-approximation} and \Eq\eqref{eq:drondphi-i} one obtains expressions for the two imposed conditions and can translate them into constraints on $\gaun$ and $\gade$. These bounds are illustrated in \fig\ref{fig:self-consistency-constraints}.  For $\gade \gg \gaun$ these conditions require
\begin{align}
\label{eq:backreaction-large-g2}
    &\frac{H^2}{2\ep \planckmass^2} \frac{\bigO(1)}{6\pi^3}e^{\pi\gade}\gatr^4\left[\frac{\gaun}{\gade}\frac{\gatr^{-\gaun}}{\gade^{\gaun}}  +\frac{\gatr^{(1-\gaun-|1-\gaun|)/2}}{\gade^{\min(1,\gaun)}} \right]\ll 1 \\
   \text{and}\quad  &\left(\frac{H}{\planckmass}
    \right)^2\frac{\bigO(1)}{6\pi^3}e^{\pi\gade}\gatr^4\frac{\gatr^{-\gaun}}{\gade^{1+\gaun}} \ll 1 \mcomma
\end{align}
where we have absorbed all the prefactors depending solely on $\gaun$ in the $\bigO(1)$ -- assuming $\gaun$ is of order unity -- and taken $\bm{\E}^2\gg \bm{\B}^2$. We have also introduced the conventional inflationary slow-roll parameter 
\begin{equation}
    \ep\equiv \dot{\phi}^2/(2\planckmass^2 H^2) \mperiod
\end{equation}
Due to the less blue spectrum, for $\ga_1>0$, we obtain weaker constraints from back-reaction than in the case of pure axion inflation, see~\cite{vonEckardstein:2025oic,Barbon:2025wjl}.
If $\gaun = 0$, these  constraints are equivalent to the ones obtained in \cite{Barnaby:2011vw}, up to powers of $\gade$ which are handled differently. As $\ep$ is small, the first constraint coming from the correction to the inflaton equation of motion, \eqref{eq:backreaction-1st-condition}, is more stringent than the second one coming from the energy density of the gauge field, \eqref{eq:backreaction-2nd-condition}.

\subsection{Electromagnetic anisotropic stress}
\label{sub:aniso}

In order to compute the secondary gravitational waves sourced by the electromagnetic field we need to extract the anisotropic part, $\Pi_{ij}$, of the field's stress-energy tensor, $T\munu$. More precisely, we need to express its unequal-time two-point function, as the expression of the GW power spectrum we will derive in \sect\ref{sub:GW-spectrum} involves the correlation $\avg{\Pi_{ij}(\ta')\Pi_{lm}(\ta'')}$ at times $\ta'\neq\ta''$.  As sub-horizon fluctuations of the source field are still in their vacuum state, they do not source gravitational waves; therefore, we only need to focus on the super-horizon regime. Like the analysis of Sections \ref{sub:EM-eom} and \ref{sub:Em-spectra}, this problematic is not new \cite{Caprini:2003vc}. However the method we pursue below, expressing both contributions of $\E$ and $\B$ to $\Pi_{ij}$ in a unified way, followed by the analytic computation developed in Appendix \ref{asub:GW-comput}, differs from usual approach found in related works.

\subsubsection{Unequal time spectra involving \intitle{$\B$ and $\E$}}
\label{subsub:aniso-generalization}

In Fourier space, the spatial part of $T\munu$ reads in Heaviside-Lorentz units~\cite{Jackson:1975}
\bea
    T_{ij}(k) &=& \kint[p] \left[ \B_i(p) \B_j^*(p-k) + \E_i(p) \E_j^*(p-k) \right. \nonumber \\
    &&  \left. \hspace{2.2cm }-\frac{1}{2}\left(\B_l(p) \B^{l*}(p-k)+ \E_l(p) \E^{l*}(p-k) \right)\de_{ij}\right] \mcomma
 \label{eq:def-Tij}
\eea
where we recall $\B = \sqrt{1+i_1}B$ and $\E = \sqrt{1+i_1} E$. The transverse-traceless part of the stress tensor is given by
\begin{equation}
\label{eq:def-Piij}
    \Pi_{ij} =\left(p_{il}p_{jm}-\pref{2}p_{ij}p_{lm}\right)T_{lm} \mcomma
\end{equation}
where $p_{ij}(k) = \de_{ij}-\hat{k}_i\hat{k}_j$ with $\hat{\kb} = \flaf{\kb}{|\kb|}$ is the projector onto the transverse plane orthogonal to $\kb$. Note that we are using the Kronecker $\de_{ij}$ to raise and lower indices, so that $\Pi_{ij}$ appropriately scales as $(1+i_1)a^{-4}$. $\Pi_{ij}$ can be decomposed further into the $\pm$ helicity basis $\Pi\symbs$ using the polarization tensors $e^\pm_{ij}$ (see \appx\ref{a:conv}). It was shown in \cite{Caprini:2003vc} that the symmetric and anti-symmetric parts of the equal-time correlation spectra for $\Pi$ without electric field is given by\footnote{In Ref. \cite{Caprini:2003vc} there is an extra factor $1/(4\pi)^2$ stemming from the use of Gaussian units in that paper. We are using Heaviside-Lorentz units, in order for the gauge fields to be canonically normalized. We refer the reader to the clear and concise appendix in \cite{Jackson:1975} for further details on electromagnetic units.}  
\begin{align}
\label{eq:aniso-correl-equaltime}
    S_\Pi(k,\ta=\ta')[\B,\E=0] &= \frac{2}{(4\pi)^3}\int\dd[3]p \Big[S_\B(|p|)S_\B(|k-p|)(1+\mu^2)(1+\bt^2) \nonumber \\
    &  \hspace{2.7cm}+ A_\B(|p|)A_\B(|k-p|) 4\mu\bt\Big] \\
\label{eq:aniso-correl-equaltime-anti}
   A_\Pi(k,\ta=\ta')[\B,\E=0] &= \frac{2}{(4\pi)^3}\int\dd[3]p S_\B(|p|)A_\B(|k-p|)4(1+\mu^2)\bt \mcomma
\end{align}

where $\mu = \hat{\bm{k}}\cdot\hat{\bm{p}}$, $\bt = \hat{\bm{k}}\cdot\widehat{\bm{k}-\bm{p}}$, and all $S_\B$'s and $A_\B$'s are evaluated at {\it equal times} $\ta=\ta'$. For our purposes, we now generalize these expressions to account for {\it (i)} a non-vanishing electric field and {\it (ii)} unequal times. 

{\it (i)} To include the electric field contribution, we argue as follows. First, we remark that \Eqs\eqref{eq:Bk-approximation} and \eqref{eq:Ek-approximation} imply one can write 
\begin{equation}
    \label{eq:little-formula-E-B}
\E\symbs = \th^\pm(\tau) f(k)\B\symbs\mcomma
\end{equation}
 where
$\th^\pm(\ta) \equiv \pm2i(2i\ta)^{|1-\gaun|/2 -(1+\gaun)/2}\times\de^\pm_E(\gaun,\gade)/\de^\pm_B(\gaun,\gade)
$ does not depend on $k$, $f(k)=k^{|1-\gaun|/2 -(1+\gaun)/2}$ does not depend on time or polarization, and $\theta \equiv \left|\theta^+\right| = \left|\theta^-\right|$ depends neither on $k$ nor on polarization. Then, we replace $\E_i(p)$  in \Eq\eqref{eq:def-Tij} by \Eq\eqref{eq:little-formula-E-B}, using the expression of $\E_i$ and $\B_i$ in the polarization basis $(\hat{k},\eps_k^+,\eps_k^-)$. The function $|\theta^\pm(\tau)|^2$ factorizes out of the integral, and we obtain the concise expression
\begin{equation}
    \Pi_{ij}[\B,\E] = \Pi_{ij}[\B,0]+\th^2 \Pi_{ij}[f\B,0]\mperiod
\end{equation}
where by $f\B$ we mean that one must replace $\B(p)$ by $f(p)\B(p)$ in the definition \eqref{eq:def-Tij}. A tensor computation similar to the \appx A of~\cite{Caprini:2003vc} then shows 
\begin{align}
\label{eq:aniso-correl-both-B-E}
        S_\Pi(k,\ta=\ta')[\B,\E] &=  S_\Pi(k,\ta=\ta')[\B,0]\nonumber \\
        &+ 2\th^2(\tau) S_\Pi(k,\ta=\ta')[\sqrt{f}\B, 0]\nonumber \\
        &+\th^4(\ta)S_\Pi(k,\ta=\ta')[f\B,0] \mperiod
\end{align}

{\it (ii)} To incorporate also the correlation at unequal times (first without electric field), we again follow closely the derivation of \Eqs\eqref{eq:aniso-correl-equaltime}, \eqref{eq:aniso-correl-equaltime-anti} presented in \cite{Caprini:2003vc}, which relies heavily on Wick's theorem, i.e. on the Gaussianity of the gauge field. As we have considered linearly evolving fields, Wick's theorem remains applicable to products evaluated at different times\footnote{One can always write $\B(\ta) = T(\ta,\ta_0)\B(\ta_0)$ with $T$ some deterministic transfer function and $\B(\ta_0)$ being an initial Gaussian field by assumption. Any product $\avg{\B(\ta_1)\cdots \B(\ta_n)}$ is then naturally related to $\avg{\B(\ta_0)^n}$, on which Wick's theorem applies.}. Moreover, using $\avg{\B_i(\kb,\tau)\B_j(\kb',\ta')} = \pref{2}(2\pi)^3\de^{(3)}(\kb-\kb')[p_{ij}S_\B(k,\ta,\ta') + i\epsilon_{ijl}\hat{k}_l A_\B(k,\ta,\ta')]$, we conclude in a very similar manner that \Eq\eqref{eq:aniso-correl-equaltime} still holds, simply by evaluating $S_\B$ and $A_\B$ at unequal times $\ta\neq\ta'$, i.e. $S_\B(k,\ta,\ta')$ and $A_\B(k,\ta,\ta')$.

\smallskip
We now combine {\it (i)} and {\it (ii)}. Let us note that $\ta\neq\ta'$ also affects the terms of \Eq\eqref{eq:aniso-correl-both-B-E}, the coefficients $2\th^2(\ta)$ and $\theta^4(\ta)$ being respectively replaced by $\th^2(\ta)+\th^2(\ta')$ and $\th^2(\ta)\th^2(\ta')$. With this, we finally obtain the generalization of \Eqs\eqref{eq:aniso-correl-equaltime} and \eqref{eq:aniso-correl-equaltime-anti}
\begin{align}
    \label{eq:aniso-correl-final}
    S_\Pi(k,\ta,\ta')[\B,\E]  &=  S_\Pi(k,\ta,\ta')[\B,0]\nonumber \\
        &+ (\th^2(\tau)+\th^2(\ta')) S_\Pi(k,\ta,\ta')[\sqrt{f}\B, 0]\nonumber \\
        &+\th^2(\ta)\th^2(\ta')S_\Pi(k,\ta,\ta')[f\B,0]\mcomma
\end{align}
where 
\begin{align}
\label{eq:angular-integral}
 \nonumber   S_\Pi(k,\ta,\ta')[\zeta\B,0] = \frac{2}{(4\pi)^3}\int\dd[3]p &\left[S_{\zeta\B}(p, \ta,\ta')S_{\zeta\B}(|k-p|,\ta,\ta')(1+\ga^2)(1+\bt^2)\right. \\
 &\left.+ A_{\zeta\B}(p,\ta,\ta')A_{\zeta\B}(|k-p|,\ta,\ta') 4\ga\bt\right]\mperiod
\end{align}

$S_{\zeta\B}$, $A_{\zeta\B}$ for $\zeta(p) \in \{1,p^{|1-\gaun|/4 -(1+\gaun)/4},p^{|1-\gaun|/2 -(1+\gaun)/2}\}$ can be obtained from \eqref{eq:S-Bonly} and \eqref{eq:A-EB}. A similar formula can   be written for $A_\Pi$. Note that this procedure only works since the time and $k$-dependence factor out in the power spectra of $\B$ and $\E$ which happens in the power law approximation.  It would not work for the more complicated exact expressions for the gauge field.

\subsubsection{UV-cutoff}
\label{subsub:aniso-cutoff}

Let us now evaluate the integral \eqref{eq:angular-integral} using our approximations for the electric and magnetic field mode functions. Because the electromagnetic field spectra are typically  blue, this integral naturally exhibits a UV divergence, which we handle with the introduction of a UV cutoff $\Lambda$. As is commonly done in the literature \cite{Sobol:2020lec,Durrer:2023rhc}, this cutoff can be set to the scale of the time-dependent ``electromagnetic horizon" defined in \Eq\eqref{eq:def-gamma3-with-min}, $\Lambda(\ta)=k_h(\ta)$. Indeed, the gauge field  modes with $k > k_h$ are still very close to their vacuum state, and therefore do not contribute to the gravitational waves source term that we aim to calculate. Furthermore, when two different times $\ta\neq\ta'$ are involved, we shall set $\Lambda$ to the lowest possible energy scale (i.e. the largest horizon), in order to guarantee that all included modes $k < \Lambda$ are super-horizon at {\it both} times $\ta$ {\it and} $\ta'$. This refines the definition \eqref{eq:def-gamma3-with-min} to 
\begin{equation}
    \label{eq:cutoff-choice}
    \La(\ta,\ta') \equiv \min(k_h(\ta),k_h(\ta')) = 
    \frac{\ga_3}{\max(-\ta,-\ta')} > 0 \mcomma
\end{equation}
Let us emphasize that for $\gaun=\gade=0$, $\La$ evaluates to zero, as it should. Indeed, without the presence of the couplings, conformal invariance is recovered and the gauge field remains in the Minkowski vacuum for {\it all} modes, so that no gravitational waves are sourced. With this procedure we mimic what would normally require a proper renormalization scheme in order not to generate gravitational waves from an electromagnetic field in its vacuum state. 

For $k\ll \Lambda$ we approximate the integration volume of  \eqref{eq:angular-integral} given by $\{\bm{p}, |\bm{p}|<\La $ and $ |\bm{k}-\bm{p}|<\La\}$ simply by $\{\bm{p},|\bm{p}|<\Lambda\}$. Splitting the integral into the contributions from $p<k$ and $p>k$, we derive an exact, fully analytic expression involving power series of $k/\Lambda$ which is detailed in \appx\ref{asub:GW-comput}. Although we will use this analytic expression in what follows, we here provide the leading order in $\Lambda$, 
\begin{align}
    \label{eq:angular-integral-symS}
  &\frac{2}{(4\pi)^3}\int_{|p|<\La} |p|^{1+s} |k-p|^{1+s}(1+\mu^2)(1+\bt^2)\dd[3]p \underset{k\ll \La}{=} \frac{1}{(4\pi)^2}\frac{56}{15(5+2s)}\La^{5+2s} + \bigO(k^2 \La^{3+2s}) \\
  \label{eq:angular-integral-symA}
  & \frac{2}{(4\pi)^3}\int_{|p|<\La} |p|^{1+a} |k-p|^{1+a}4\mu\bt\dd[3]p \underset{k\ll \La}{=} -\frac{1}{(4\pi)^2}\frac{8}{3(5+2a)}\La^{5+2a} + \bigO(k^2\La^{3+2a}) \\
  \label{eq:angular-integral-antisym}
   &\frac{2}{(4\pi)^3}\int_{|p|<\La} |p|^{1+s}|k-p|^{1+a}4(1+\mu^2)\bt \dd[3]p \underset{k\ll \La}{=} \frac{1}{(4\pi)^2}\frac{32(5+2a)}{15(4+s+a)}k\La^{4+s+a} + \bigO(k^3 \La^{2+s+a}) \mperiod
\end{align}
for any $s,a\in\mathbb{R}$ such that, respectively, $5+2s>0$, $5+2a>0$ and $4+s+a>0$. In the case where e.g. $5+2s<0$, the corresponding integral is actually led by $k^{5+2s}$ rather than $\La^{5+2s}$, however no short expression exists because all orders in the power series contribute to such a term, and we defer the reader to \appx\ref{asub:GW-comput}. Finally, although expression \eqref{eq:angular-integral-symS} (similar considerations apply to \eqref{eq:angular-integral-symA} and \eqref{eq:angular-integral-antisym}) may appear divergent at $5+2s=0$, its complete expression is not, as powers of $\La$ and $k$ typically cancel like $\flaf{(\La^{5+2s}-k^{5+2s})}{(5+2s)} \underset{5+2s\to 0}{\longrightarrow} \ln(\La/k)$. Quite importantly, the result for $5+2s>0$ is a white noise depending solely on $\La$, as expected from the integration of a blue spectrum up to a UV-cutoff.  For $5+2s< 0$ instead, $S_\Pi\propto k^{5+2s}$ becomes red.  In \Eq\eqref{eq:aniso-correl-final}, we see that this change of behavior from white noise for small $\gaun$ to a red spectrum happens to $S_\Pi$ (resp. $A_\Pi$) at $\gaun = 5/2$ (resp. $\gaun = 2$) for the electric contribution, at $\gaun = 9/2$ (resp. $\gaun= 4$) for the magnetic contribution and at $\gaun= 7/2$ (resp. $\gaun= 3$) in the cross-term. We also note that $A_\Pi\propto k \La^{4+s+a}$ rather than $\propto \La^{5+s+a}$ for $4+s+a>0$ because  the associated angular integral vanishes in the limit $k\ra 0$.

 Ultimately, we obtain
\begin{align}
     \nonumber
     S_\Pi(k,\ta,\ta')
     =& \pref{(4\pi)^2}\frac{1}{4a^4(\ta)a^4(\ta')}\times\left[\vphantom{\frac{\pi\gade}{3}}\right. \\
     \nonumber
     &\De_B^2 (\ta\ta')^{m_B} \times\left(\cosh^2\left(\frac{\pi\gade}{2}\right) f_S(\La,k,m_B) + \sinh^2\left(\frac{\pi\gade}{2}\right) f_A(\La, k, m_B)\right) 
   \\ 
  \nonumber & + \De_E^2 (\ta\ta')^{m_E} \times \{m_B \longrightarrow m_E\} \\ 
  & +\left. \De_B\De_E (|\ta|^{m_B}|\ta'|^{m_E} + |\ta|^{m_E}|\ta'|^{m_B})\times \{m_B \longrightarrow \flaf{(m_B+m_E)}{2}\}
     \vphantom{\frac{\pi\gade}{3}}\right]\mcomma
       \label{eq:final-S-Pi}
\end{align}
\begin{align}
     \nonumber
     A_\Pi(k,\ta,\ta')
     =& \pref{(4\pi)^2}\frac{1}{4a^4(\ta)a^4(\ta')}\frac{\sinh(\pi\gade)}{2}\times\left[\vphantom{\frac{\pi\gade}{3}}\right. \\
     \nonumber
     &\De_B^2 (\ta\ta')^{m_B}  g(\La,k,m_B)  
   + \De_E^2 (\ta\ta')^{m_E} g(\La,k,m_E)   \\ 
  +& \left. \De_B\De_E (|\ta|^{m_B}|\ta'|^{m_E} + |\ta|^{m_E}|\ta'|^{m_B})g(\La,k,\flaf{(m_B+m_E)}{2})  
     \vphantom{\frac{\pi\gade}{3}}\right]\mcomma
       \label{eq:final-A-Pi}
\end{align}
where
\begin{equation}
    \label{def:mB-mE}
    m_B \equiv 1-|1-\gaun| \spliteq m_E\equiv -\gaun \mperiod
\end{equation}
The functions $f_S$, $f_A$, $g$ are provided in \appx\ref{asub:GW-comput}. Based on the previous paragraph, depending on the value of their last argument, they tend to be dominated either by $\La$ or by $k$. We recall that $\La=\La(\ta,\ta')$ is time-dependent and given by \Eq\eqref{eq:cutoff-choice}.

In each square bracket above, the first (resp. second) term comes from the contribution of the sole magnetic (resp. electric field), while the third is the mixed term where both fields contribute. We find that $S_\Pi$ is non-divergent for all $0\leqslant\gaun < 4$, in particular at $\gaun\in\{5/2,7/2\}$. In the regime where $S_\Pi,A_\Pi$ are white noise, the dimensionless\footnote{C.f. footnote \ref{foot:dimensionless}.} power spectra per logarithmic interval $\PP_\Pi$ and $\PP^A_\Pi$ thus scale as $\propto k^3$.

In \appx\ref{asub:GW-comput} we compare our analytical expressions for $S_\Pi,A_\Pi$ against a numerical integration and show that they are in satisfactory agreement for $k|\ta|\ll  1$. We have also numerically asserted that the leading terms in the formulae \eqref{eq:angular-integral-symS}--\eqref{eq:angular-integral-antisym} are in excellent agreement with the value of the whole integral.

\section{Gravitational waves induced by  gauge fields}\label{s:GWprod}

Gauge fields have a transverse-traceless contribution to the stress tensor that has been computed in the last section. Even though the coherence scale (wavelength) of the physical modes is super-horizon, it will induce a transverse-traceless perturbation to the metric. During inflation this will simply contribute to a shear in the spacetime geometry. But once these perturbations re-enter the horizon during the radiation or matter dominated era, they will start oscillating and behave as normal gravitational waves.
In this section we calculate the power spectrum of these tensor perturbations that will turn into a power spectrum of oscillating gravitational waves in the late Universe.

\subsection{Sourced gravitational waves in a Friedmann metric}
\label{sub:GW-eom}

We consider linear  perturbation to the metric \eqref{eq:FRLW-metric} describing gravitational waves,
\begin{equation}
    g\munu = a^2(\eta\munu+h\munu) \mcomma
\end{equation}
with $|h\munu|\, \ll 1$. In the transverse-traceless (TT) gauge, this perturbation is fully described by its spatial part $h_{ij}$, satisfying $h_{ii}=0$ and $h_{ij}\hat{k}^j=0$ (in Fourier space). Note that, following \cite{maggiore_vol2}, we are moving indices of $h_{ij}$ with the Kronecker $\de_{ij}$. Separating the  polarizations into its two helicities $h\symbs$ (c.f. \appx\ref{a:conv}), the linearized Einstein equations reduce to the well-known propagation equation for tensor perturbations of the Friedmann metric \cite{maggiore_vol2},
 \begin{equation}
      \label{eq:GWeq-hvariable}
    h_k^{\pm \prime\prime} +2 \HH h_k^{\pm \prime} +k^2 h\symbs = 16\pi G a^2 \Pi\symbs \mcomma
 \end{equation}
with $\Pi\symbs$ the two helicities of the TT stress tensor $\Pi_{ij}$ defined in \sect\ref{sub:aniso} (again, c.f. \appx\ref{a:conv}). This equation can be re-written using the comoving metric perturbation $\chi_{ij}(k,\ta) = a(\ta)h_{ij}(k,\ta)$, yielding
\begin{equation}
       \label{eq:GWeq}
    \chi_k^{\pm \prime\prime} +\left(k^2 - \frac{a''}{a}\right)\chi\symbs = 16\pi G a^3 \Pi\symbs \mperiod
\end{equation}
As discussed in \sect\ref{s:electromag}, within standard electromagnetism the $\Pi\symbs$ in the above equations scales as $\propto a^{-4}$. 
Introducing $x\equiv k\tau$ we can write \eqref{eq:GWeq} in the form
\be   \label{eq:GWeq2}
\dv[2]{\chi\symbs}{x} +\left(1-\pref{a}\dv[2]{a}{x}\right)\chi\symbs = \frac{16\pi G a^3}{k^2} \Pi\symbs \mperiod
\ee
The solution of \Eq\eqref{eq:GWeq2} is expressed in terms of the retarded Green function $G(x,y)$ of that same equation, by
\begin{align}
\label{eq:solution-GW-eq-h}
   h\symbs(\ta) &= \pref{a(\ta)} \chi\symbs(\ta) = \frac{16\pi G}{a k^2}\int^{x}_{x_i} G(x,y) a^3(y)\Pi\symbs(y)\dd y  \\
       \label{eq:solution-GW-eq-hprime}
   h_k^{\pm \prime}(\ta) &=  \frac{16\pi G}{a k} \int_{x_i}^x G_1(x,y)a^3(y)\Pi\symbs(y)\dd y \mcomma
\end{align}
where $x_i=k\ta_i$ is the time at  which the source starts generating gravitational waves. We postpone the precise definition of $x_i$ to the end of this section. We have also defined $G_1(x,y) \equiv \left(\pdv{}{x} - \frac{1}{a}\dv{a}{x}\right) G(x,y) $ (and we have used $G(x,x)=0$). The Green function has the general form
\begin{equation}
\label{eq:theory-green-function-formula}
    G(x,y) = \pref{W(y)}[u_1(y)u_2(x)-u_2(y)u_1(x)]\Theta(x-y) \mcomma
\end{equation}
where $\Theta$ is the Heaviside step function, and $u_1,u_2$ are any two independent homogeneous solutions of \Eq\eqref{eq:GWeq2} with Wronskian $W(y) \equiv u_1 \partial_y u_2 - u_2\partial_y u_1$.

For a generic power law expansion, $a\propto \ta^{2/(1+3w)}$ with a constant $w$, a possible set of homogeneous solutions is 
\begin{equation}
    u_1(k\ta) = k\ta\, j_{\nu(w)}(k\ta)  \spliteq u_2(k\ta) = k\ta\, y_{\nu(w)} (k\ta)  \mcomma
\end{equation}
with 
\begin{equation}
    \nu(w) = \frac{1-3w}{1+3w} =\frac{2}{1+3w}-1 \mperiod
\end{equation}
Here $j_\nu$ and $y_\nu$ are the spherical Bessel functions, and the Wronskian of these solutions is 1. 
In the special case $w=-1$ (lowest order slow-roll approximation, $a''/a = 2/\ta^2$), $\nu=-2$ and we are led to 
\begin{equation}
\label{eq:Green-from-sin-and-cos}
    G(x,y) =  \frac{(1+xy)\sin(x-y)-(x-y)\cos(x-y)}{xy}\Theta(x-y)\mperiod
\end{equation}
This is the Green function for $w=-1$. Including the slow-roll parameter $\ep$ to lowest order (i.e. $\ep \simeq \const$ but non zero) such that $a =(-H\tau)^{-1-\ep}$, one finds $w=-1+2\ep/3$ and $\nu=-2-\ep$. This manifests itself in a small tilt of the final GW power spectrum, in a similar fashion as for standard slow-roll tensor perturbations \cite{maggiore_vol2}. In this analytical derivation we first neglect these slow-roll corrections, and explain briefly at the end how to reincorporate them.

The initial value $x_i = k\ta_i$ at which the source is turned on is the time at which the electromagnetic source of scale $1/k$ exits its horizon, which we have defined in \Eq\eqref{eq:def-electro-horizon-crossing-time}: $x_i = k\ta_h(k) = -\gatr$. Indeed, fields in their vacuum state do not contribute to the energy budget of the Universe, hence cannot source gravitational waves. Let us recall that the electromagnetic field at scale $1/k$ exits its vacuum state when it crosses the pseudo-horizon we have studied in \sect\ref{s:electromag}, not the regular Hubble horizon. This pseudo-horizon depends on the coupling parameters $\gaun, \gade$ and is pushed to infinity when $\gaun=\gade=0$, so that no gravitational waves are sourced if no gauge fields are amplified out of the Bunch-Davies vacuum.

Furthermore, the solution \eqref{eq:solution-GW-eq-h} we have chosen does not feature a homogeneous part, hence $h\symbs = 0$ at $\ta_i=\ta_h$. In principle, one should rather consider the tensor perturbations to start in their vacuum state as well, therefore using Bunch-Davies initial conditions for $h\symbs$ instead of zero. This gives rise to the well-studied primordial tensor perturbations from inflation~\cite{maggiore_vol2}. However, these primordial gravitational waves are uncorrelated to the secondary perturbations we are considering here, since they are generated from the vacuum fluctuations of different, independent fields. Consequently, one can always separate the contributions to any quadratic expectation value from both types of waves without having to worry about cross-correlations, i.e. $\PP\subsc{tot} = \PP\subsc{vacuum} + \PP\subsc{sourced}$. We can therefore safely focus on the sourced  contribution and ignore the primordial one that has been thoroughly studied in the literature.

\subsection{Power spectrum of gravitational waves and their derivatives during inflation}
\label{sub:GW-spectrum}

From the solution \Eq\eqref{eq:solution-GW-eq-h}, we aim at describing the power spectrum of the induced gravitational waves during inflation, as a first step towards computing them in later cosmological eras. The symmetric and anti-symmetric spectra $S_h$, $A_h$, $S_{h'}$ and $A_{h'}$ for $h$ and its derivative are defined using the general \Eq\eqref{eq:def-unequaltime-correlator-general} with appropriate normalization of the polarization tensors. The GW power spectrum and energy density per logarithmic interval of $k$ are then defined by~\cite{Caprini:2018mtu}
\begin{align}
\label{eq:def-GW-curl-power-spectrum}
    \PP_T(k,\ta) &= \PP_h(k,\ta) = \frac{k^3}{2\pi^2} S_h(k,\ta,\ta)
    \\
\label{eq:def-GW-curl-antisym-power-spectrum}
    \PP_T^A(k,\ta) &= \PP_h^A(k,\ta) = \frac{k^3}{2\pi^2} A_h(k,\ta,\ta)
\\
\label{eq:def-GW-energy-density}
    \dv{\rho_{gw}}{\log k}{} (k,\ta) &= \pref{32\pi G a^2} \frac{k^3}{2\pi^2} S_{h'}(k,\ta,\ta)\mperiod
\end{align}
Of course $\rho\ssgw$ cannot be truly interpreted as an `energy density' as long as the tensor perturbations have not reentered the horizon. Nevertheless, the mathematical definitions being similar, we use the same notations. It is also useful to define 
\begin{equation}
\label{eq:def-Omega_GW}
    \Om\ssgw(k,\tau) \equiv \frac{1}{\rho_c}\dv{\rho\ssgw}{\log k} =\frac{k^3}{24\pi^2\HH^2} S_{h'}(k,\ta,\ta) \mcomma
\end{equation}
where $\rho_c(t) = 3\planckmass^2 H^2=3\HH^2/(8\pi Ga^2)$ is the critical density. We also define the antisymmetric counterpart of \eqref{eq:def-Omega_GW},
\begin{equation}
\label{eq:def-Omega_AGW}
    \Om\ssgw^A(k,\tau) =\frac{k^3}{24\pi^2\HH^2} A_{h'}(k,\ta,\ta) \mperiod
\end{equation}
Inserting the solutions~\eqref{eq:solution-GW-eq-h} and \eqref{eq:solution-GW-eq-hprime}, we can relate these quantities to the properties of the source,
\begin{align}
\label{eq:PT-from-double-int}
    \PP_T(k,\tau) &= \frac{2}{\pi^2 \planckmass^4}\frac{1}{k a^2}\int^{x}_{-\gatr}\! \int^{x}_{-\gatr}\!\! G(x,y)G(x,z) a^3(y)a^3(z)S_\Pi(k,k^{-1}y,k^{-1}z)\dd y \dd z \\
\label{eq:Omega-from-double-int}
    \Om\ssgw(k,\tau) &= \frac{1}{6\pi^2 \planckmass^4}\frac{k}{H^2 a^4}\int^{x}_{-\gatr}\! \int^{x}_{-\gatr}\!\! G_1(x,y)G_1(x,z) a^3(y)a^3(z)S_\Pi(k,k^{-1}y,k^{-1}z)\dd y \dd z \mperiod
\end{align}
Moreover, $\PP_T^A$ and $\Om\ssgw^A(k,\tau)$ are given upon the replacement of $S_\Pi$ by $A_\Pi$ in the above equations.

To compute these physical quantities we therefore need the {\it unequal time} correlations of $\Pi_{ij}$, as the above double integrals involve correlating $\avg{\Pi\symbs(\ta')\Pi\symbs(\ta'')}$ between any two instants $\ta_h(k) \leqslant \ta',\ta'' \leqslant \ta$.  In Section \sect\ref{s:electromag}, we have built the tools required to perform such a computation. Crucially, the UV cutoff appearing in the expression for $S_\Pi$ (see \Eq\eqref{eq:final-S-Pi})  is a function of both times $\ta'$ and $\ta''$, ensuring that no vacuum field is accounted for as a GW source. We 
refer the reader to \appx\ref{asub:GW-comput} for further details on the assumptions required in the analytic computation of the integrals~\eqref{eq:PT-from-double-int}--\eqref{eq:Omega-from-double-int}. The final expressions in the superhorizon limit $|k\ta|\ll \gatr$ are
\begin{align}
    \label{eq:final-PT}
   \PP_T(k,\tau) &= \frac{H^4}{\planckmass^4} \FF(\ga_1,\ga_2) \\
   \label{eq:finalPAT}
      \PP^A_T(k,\tau) &= \frac{H^4}{\planckmass^4} \FF_A(\ga_1,\ga_2) \\
    \label{eq:finalOmGW}
    \Om\ssgw(k,\ta) &=  \frac{H^4}{\planckmass^4}(-k\ta)^3 \FF'(\ga_1,\ga_2,k\ta) \\
    \label{eq:finalOmAGW}
   \Om^A\ssgw(k,\ta) &= \frac{H^4}{\planckmass^4}(-k\ta)^4 \FF'_A(\ga_1,\ga_2)\mperiod
\end{align}
The complicated functions $\FF,\FF_A,\FF',\FF'_A$ typically depend exponentially on $\ga_2$ but more weakly on $\ga_1$, unless $\gaun\to 4$ (due to the infrared divergence of the electric field). Their full expressions, which are well-defined and finite for any $(\gaun,\gade)\in \interval{[}{0}{4}{[}\cross\mathbb{R}$ are presented and discussed in Appendix~\ref{asub:GW-comput}. The residual dependence of $\FF'$ on $k\ta$ is very weak compared to the prefactor $(-k\ta)^3$. As an example, we show here the pure electric field contribution to $\PP_T$,
\begin{equation}
\left.\PP_T\right|_{\text{pure}\, E}
     = \frac{H^4}{\planckmass^4} \frac{\gatr^8}{2\pi^2(4\pi)^2}\left[ 
      \frac{\De_E^2 \gatr^{-2 \gaun}}{9(4-\gaun)^2}\left(A_1\cosh^2\left(\frac{\pi\gade}{2}\right) + A_2\sinh^2\left(\frac{\pi\gade}{2}\right)\right)\right] \mcomma
       \label{eq:PT_E-only}
\end{equation}
where $A_1$, $A_2$ are slowly-varying functions of $\gaun$. The exponential dependence of $\FF$ on $\ga_2$ stems from the hyperbolic functions as well as from $\De_E$, which also grows exponentially for large  $\gade$.

\begin{figure}
    \centering

\includegraphics[width = 0.6\linewidth]{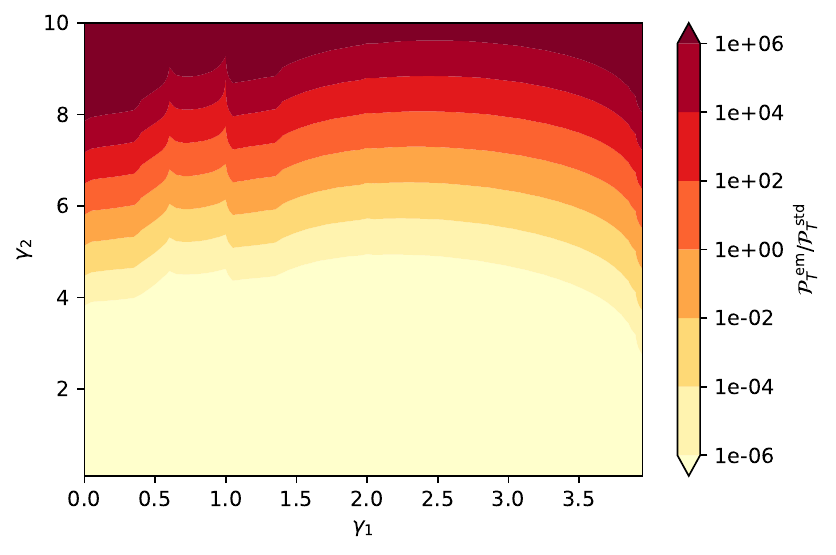}
\includegraphics[width=0.6\linewidth]{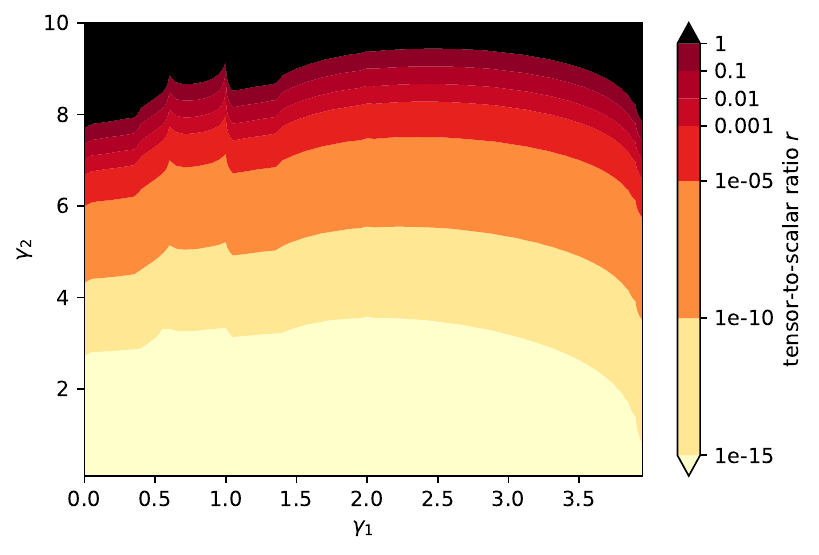}
   
    \caption{{\it Upper panel:} Comparison of prediction \eqref{eq:final-PT} to the vacuum inflationary tensor power spectrum $\PP_T = 2H^2/(\pi^2 \planckmass^2)$ (see \Eq\eqref{eq:PT-std-inflation}). {\it Lower panel:} The contribution of gauge fields induced GWs to the tensor-to-scalar ratio $r(\ga_1,\ga_2)$ for different parameters $\gaun,\gade$. This ratio is computed using our \Eq\eqref{eq:final-PT} for the tensor amplitude, and comparing it to the scalar power spectrum measured by Planck (details are given in Section~\ref{s:GWres}). That is, we assume the scalar amplitude to be known from experiment and to avoid complications due to spectral tilts we have set $k$ to the tensor pivot scale, $k=k_* =  \SI{0.002}{\mega\mathrm{pc}^{-1}}$, so more explicitely $\PP_{\mathcal{R}}(k_*) = \SI{2.4e-9}{}$. The `spike' at $\ga_1=1$ is due to a peak in the magnetic field power spectrum that dominates in this region, while the spike at $\ga_1=0.4$ comes from the electric field, that is somewhat poorly modeled around $\ga_1=0.4$, see \fig\ref{fig:electro-fields}. Near $\gaun =4$ the electric field is diverging, which increases $r$ quite drastically. In both panels we have taken $H\ssend \simeq \SI{2.2e12}{\GeV}$, see below.}
    \label{fig:powerspectrum}
\end{figure}

A striking consequence of this calculation is that the gravitational wave power spectrum $\PP_T$ is scale invariant. Some of its values are illustrated in \fig\ref{fig:powerspectrum}. Including the slow-roll corrections to the scale factor and the Green function at first order, this power spectrum becomes tilted by a factor $(k/k_*)^{n_T}$ with $n_T=-4\ep$ and $k_*$ some arbitrary pivot scale, hence its tilt differs from standard inflationary tensor perturbations \cite{Teuscher:2025xke}.

\subsection{Evolution after inflation}
\label{sub:other-eras}
Next we compute the spectrum of the GWs generated during inflation, after they have evolved through later cosmological eras. We assume that the source shuts off at the end of slow-roll inflation $\ta\ssend$, so GWs evolve freely after $\ta\ssend$. This is not exactly correct. Only the electric field is exponentially damped by the presence of charges, but the magnetic field especially on large scales will survive. However, on small scales the magnetic field is damped by diffusion, and since the gauge field power spectrum is blue and the anisotropic stress is mostly dominated by the contribution from the electric field, neglecting the source after inflation is a good approximation. 

To continue, we  match the free solution to the initial conditions provided by inflation. For the sake of generality, and for the reader interested in including more exotic periods of expansion than the radiation and matter dominated eras in the cosmological history, we derive here a general and efficient method to perform this matching for an arbitrary number of successive (constant) equations of state. This can help e.g. to incorporate the effect of reheating, which may exhibit e.g. a kination dominated phase~\cite{Gouttenoire:2021jhk}. We will then write explicitly the results for the standard cosmological scenario.

\subsubsection{The general case}

Let us start by supposing the post-inflationary Universe to be described by a succession of eras with equations of state $w_1,\dots,w_n$ that instantaneously switch from one to the next at times $\ta_{1|2},\dots,\ta_{n-1|n}$. That is to say,
the scale factor has the piecewise expression
\bea
    a(\ta) &=& a_{i-1|i}\left[\frac{1+3w_i}{2}\HH_{i-1|i}(\ta-\ta_{i-1|i})+1\right]^{2/(1+3w_i)} \nonumber \\
    &=& a_{i|i+1}\left[\frac{1+3w_i}{2}\HH_{i|i+1}(\ta-\ta_{i|i+1})+1\right]^{2/(1+3w_i)}\mcomma
\eea
during the period $\ta_{i-1|i} \leqslant \ta \leqslant \ta_{i|i+1}$, with $\HH_{i|i+1} = \HH(\ta_{i|i+1})$ and $a_{i|i+1}=a(\ta_{i|i+1})$. The first (resp. second) expression is to be used when matching the $i$-th era with the previous (resp. next) one. Consequently, the solution of \Eq\eqref{eq:GWeq} with vanishing source term during the $i$-th era takes the form  (we drop the polarization labels)
\begin{eqnarray}
    a(\ta)h_k(\ta) =\chi_k(\ta) &=& A_k^i\, k\tita_i\, j_{\nu_i}(k\tita_i) + B_k^i\, k\tita_i\, y_{\nu_i}(k\tita_i)\mcomma \\
    \nu_i &\equiv& -1+\flaf{2}{(1+3w_i)} \mcomma \\
  \tita_i &= &\ta -\ta_{i-1|i}+(\nu_i+1)\HH_{i-1|i}^{-1} = \ta -\ta_{i|i+1}+(\nu_i+1)\HH_{i|i+1}^{-1}\mcomma
\end{eqnarray}
where we recall that $j_\nu,y_\nu$ are the spherical Bessel functions (see Section \ref{sub:GW-eom}).  

We assume the continuity of the scale factor and Hubble parameter (the first and second fundamental forms) at the transition. Using properties of Bessel functions \cite{Abramo}, matching the GW amplitude and its time derivative across each transition leads to the following relations between the coefficients $A_k^i, B_k^i$ and $A_k^{i+1}, B_k^{i+1}$:
\begin{equation}
    M_i[\ell_i] \begin{pmatrix}
       A_k^i \\
       B_k^i
    \end{pmatrix}
  =  M_{i+1}[r_{i+1}] \begin{pmatrix}
       A_k^{i+1} \\
       B_k^{i+1}
    \end{pmatrix}\mcomma
\end{equation}
where
\begin{align}
\label{eq:matrix-bessel}
      M_i[x] &\equiv 
\begin{pmatrix}
    x j_{\nu_i}(x) & x y_{\nu_i}(x) \\
    (\nu_i+1)j_{\nu_i}(x)- x j_{\nu_i+1}(x) ~~ &  ~~  (\nu_i+1)y_{\nu_i}(x)- x y_{\nu_i+1}(x) 
\end{pmatrix}\\
\label{eq:-left-of-era-transition}
\ell_i &\equiv \frac{2k}{\HH_{i|i+1} (1+3 w_i)}\quad \forall\, 1\leqslant i < n \\
\label{eq:right-of-era-transition}
r_i &\equiv \frac{2k}{\HH_{i-1|i} (1+3 w_i)} \quad \forall\, 1<i\leqslant n \mperiod
\end{align}
 This allows us to express successive  coefficients in terms of transition matrices $T_i$'s given by
 \begin{equation}
     T_i\equiv M_{i+1}[r_{i+1}]^{-1}M_i[\ell_i]\mperiod
 \end{equation}
The final coefficients are 
\begin{equation}
\label{eq:the-chain}
    \begin{pmatrix}
       A_k^{n} \\
       B_k^{n}
    \end{pmatrix}=T_{n-1} \cdots T_1 \begin{pmatrix}
       A_k^{1} \\
       B_k^{1}
    \end{pmatrix}\mperiod
\end{equation}
Note that $\det M_i[x] = 1$.
As the GWs are of stochastic nature, knowing these coefficients is of interest mainly to relate them to the  statistical expectation values like $\PP_T$ and $\Om\ssgw$ both at the beginning and at the end of the evolution. If, in the $n$-th era, a GW has reentered the horizon,\footnote{Note that whether the GW is super- or sub-horizon during the intermediate eras is  unimportant. It might even exit and enter the horizon several times.} then using again the properties of the Bessel functions and adapting the definition \eqref{eq:def-Omega_GW} to a single polarization yields to 
\begin{align}
\label{eq:chain-end}
\PP_{T,n}\supsc{1pol.}(k)& \simeq \frac{k^3}{2\pi^2} \frac{1}{2a^2}\left[(A_k^n)^2+(B_k^n)^2\right] \\
    \Om\supsc{1pol.}_{\text{gw},n}(k) &\simeq \pref{24\pi^2} \frac{k^5}{H^2 a^4} \pref{2}\left[(A_k^n)^2+(B_k^n)^2\right] = \frac{k^2}{12 a^2 H^2}\PP_T\supsc{1pol.}\mcomma
\end{align}
where $H$ and $a$ are to be evaluated at the appropriate time in the $n$-th era.

On the other side of the chain, the first freely propagating era is supposed to be preceded by a period during which the GW source is active (again assuming an instantaneous transition). If we denote $h(\ta\ssend)$ the amplitude of one GW polarization at the end of the source era and $\HH\ssend=\HH(\ta\ssend)$ the comoving Hubble parameter at that time, the coefficients $A_k^1, B_k^1$ are given by
\begin{equation}
\label{eq:source-to-free-chain}
      \begin{pmatrix}
       A_k^{1} \\
       B_k^{1}
    \end{pmatrix}=M_1[r_1]^{-1}\begin{pmatrix}
       a(\ta\ssend) h(\ta\ssend) \\
       (a h)'(\ta\ssend) / k
    \end{pmatrix}\mperiod
\end{equation}
Moreover, let us consider that the GWs produced by the source are very super-horizon at $\ta\ssend$. Then the GWs are not oscillating, hence the typical GW characteristics can be taken to be 
\begin{align}
    h\ssend &\simeq \sqrt{S_h\supsc{1pol.}}e(\kb) = \frac{\sqrt{2\pi^2}}{k^{3/2}}\sqrt{\PP_T\supsc{1pol.}(\ta\ssend)}e(\kb) \\
    \label{eq:chain-beginning}\qquad h'\ssend&=\pref{a\ssend}(ah)'\ssend-\HH\ssend h\ssend \simeq \sqrt{S_{h'}\supsc{1pol.}}e(\kb)= \frac{\sqrt{24\pi^2}}{k^{3/2}}\HH\ssend \sqrt{\Om\ssgw\supsc{1pol.}(\ta\ssend)}e(\kb) \mcomma
\end{align}
where $e(\kb)$ is a (in general not Gaussian) random variable with 
\be
\langle e(\kb)e^*(\kb')\rangle = \de^{(3)}(\kb-\kb') \mperiod
\ee
Therefore, combining \Eqs\eqref{eq:the-chain}--\eqref{eq:chain-beginning} links the late time observables in the $n$-th era (e.g. today) to the original GW spectrum of the source:
\begin{equation}
\label{eq:mutiple-transitions-final}
    \Om_{\text{gw},n}\supsc{1pol.}(k) = \frac{a\ssend^4 H^2\ssend}{a_n^4 H^2_n} \frac{1}{24} \left\Vert T_{n-1}\cdots T_1 M_1[r_1]^{-1} 
    \begin{pmatrix}
(\flaf{k}{\HH\ssend}) \sqrt{\PP\supsc{1pol.}_{T,\text{end}}} \\[2mm]
\sqrt{12\Om\supsc{1pol.}\subsc{gw,end}}+\sqrt{\PP\supsc{1pol.}_{T,\text{end}}}
    \end{pmatrix}\right\Vert^2 \mperiod
\end{equation}
If moreover the present time belongs to the $n$-th era, i.e. $H_n = H_0$ and $a_n=a_0$, this equation can be also be recast using $h\equiv \flaf{H_0}{\SI{100}{\kilo\meter\cdot\second^{-1}\cdot\mega\rm{pc}^{-1}}}$,
\begin{equation}
\label{eq:mutiple-transitions-indep-of-H0}
    h^2\Om_{\text{gw},0}\supsc{1pol.}(k) = \frac{2.2\times 10^{83}}{(1+z\ssend)^4}\fiducial{H\ssend}{1}{\GeV}{2}\frac{1}{24} \left\Vert T_{n-1}\cdots T_1 M_1[r_1]^{-1} 
    \begin{pmatrix}
(\flaf{k}{\HH\ssend}) \sqrt{\PP\supsc{1pol.}_{T,\text{end}}} \\[2mm]
\sqrt{12\Om\supsc{1pol.}\subsc{gw,end}}+\sqrt{\PP\supsc{1pol.}_{T,\text{end}}}
    \end{pmatrix}\right\Vert^2 \mperiod
\end{equation}

If the original source  produces different amounts of each polarization, the total energy density is  of course obtained by summing the two versions of \Eq\eqref{eq:mutiple-transitions-indep-of-H0}. This equation can therefore incorporate the effect of any number of eras with constant equations of state. The only ingredients required as inputs are the $w_i$'s, the values $\HH_{i|i+1}$ of the comoving Hubble parameter at each transition, which enter \Eqs\eqref{eq:-left-of-era-transition}--\eqref{eq:right-of-era-transition}, and the redshift $1+z_{\mathrm{end}} =a_0/a_{\mathrm{end}}$. These values can be related to more physically relevant quantities, such as the temperature and/or redshift at the transition times, using the conservation of entropy per comoving volume. Of course in reality the transitions from one expansion law to another is usually not instantaneous but gradual. However, our analytical study provides a concise and generic framework to keep control over all ingredients of the problem. A numerical, more accurate study can then naturally be performed in specific cases.

Finally, notice that the term $\sqrt{12\Om\supsc{1pol.}\subsc{gw,end}}$ accounts for the non-zero time derivative of tensor perturbations at the beginning of the radiation era. In models where perturbations are frozen on super-horizon scales (like standard inflationary tensor perturbations), this term is not present and the final energy density is a function of the initial power spectrum $\PP\supsc{1pol.}_{T,\text{end}}$ only. This does not apply to our situation where tensor perturbations are continuously sourced during inflation, even at superhorizon scale, hence their time derivative is non-vanishing.

\subsubsection{The standard cosmological scenario}

\begin{figure}
    \centering
    \includegraphics[width=0.9\linewidth]{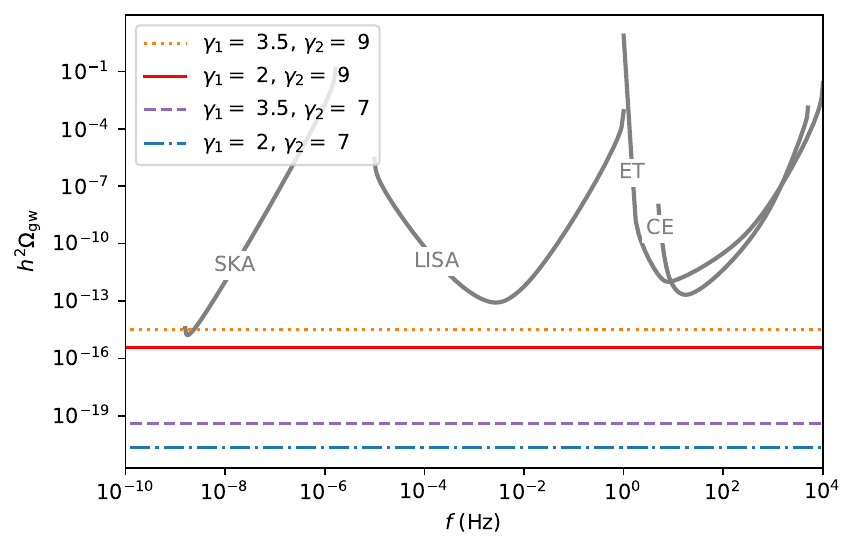}
    \caption{GW energy density per log frequency today, $f=k/(2\pi a_0)$, generated by the inflationary electromagnetic fields. Because of the Gamma and exponential functions in the expression \eqref{eq:final-PT}, the resulting energy is strongly dependent on the parameter values $\gaun,\gade$  (here $T\ssend =\SI{1e15}{\GeV}$, hence $H\ssend \simeq \SI{2.2e12}{\GeV}$). All frequencies is the displayed range reenter the horizon deep in the radiation era, resulting in a near scale invariant energy density. We also indicate sensitivity curves from different experiments in gray: `Square Kilometer Array' (SKA),  `Laser Interferometer Space Antenna' (LISA),  `Einstein Telescope' (ET) and `Cosmic Explorer' (CE).  Data are taken from \cite{alberto-roper-pol}.}
    \label{fig:omegaGW}
\end{figure}

We now apply this general method to the case of a simple transition from slow-roll inflation to a Universe filled with matter and radiation, starting with a radiation era. In particular, we neglect possible additional GW generation during reheating, see~\cite{Maiti:2025cbi} for a treatment of this question. This is certainly a good approximation for wavenumbers with $|k\ta\ssend|\ll 1$ where $\ta\ssend$ is the conformal time at the end of inflation. The expression of the scale factor in a Universe with matter and radiation is\footnote{A useful trick to account for the effect of late dark energy is to use a different normalization of $a_0$. Instead of setting $a_0=1$ (today), and  considering the effect of dark energy to raise only after e.g. redshift $z=1$, one rather sets $a_0=1+z=2$ in all the main text equations. The evolution between $z=1$ and $z=0$ can then be scrutinized separately. Here we simply neglect this correction.}
\begin{equation}
    a(\ta) = a_0^2 H_0 \ta\left(a_0 H_0 \frac{\Om_m}{4}\ta + \sqrt{\Om_r}\right)\mcomma
\end{equation}
where $\Om_m$ and $\Om_r$ denote the matter and radiation density parameters today.
From the conservation of entropy per comoving volume, one deduces the expressions 
\begin{align}
\label{eq:a-end}
a\ssend &= \frac{a_0}{\alpha\ssend}\frac{T_0}{T\ssend} \\
\label{eq:hub-end}
    \HH\ssend &\simeq \alpha\ssend \frac{T\ssend}{T_0} a_0 H_0\sqrt{\Om_r} \\
    \HH\sseq &\simeq \sqrt{2}\frac{\Om_m}{\sqrt{\Om_r}}a_0 H_0 \mcomma
\end{align}
with the `$0$' label referring to present day quantities and `eq' to the matter radiation equality. $T\ssend$ is the temperature at the end of inflation (assuming instantaneous reheating), and $\alpha\ssend \equiv \left(\frac{g_*^S(T\ssend)}{g_*^S(T_0)}\right)^{1/3} \simeq\left(\frac{106,75}{3,94}\right)^{1/3} \simeq 3$ is the variation in the number of entropic degrees of freedom \cite{Baumann:2018muz}. For the numerical value we have assumed all standard model degrees of freedom to be relativistic at the end of inflation, hence $T\ssend>\SI{100}{\GeV}$.  As mentioned earlier, together with \Eq\eqref{eq:final-PT} and \Eq\eqref{eq:finalOmGW}, these are all the ingredients needed to compute the present GW energy density \eqref{eq:mutiple-transitions-final} with $n=2$. In particular, notice that we do not need to specify in which era do the GWs reenter the horizon: this is completely accounted for in the matrix \eqref{eq:matrix-bessel} that involves Bessel functions.  Values of $h^2\Om\ssgw$ are shown in \fig\ref{fig:omegaGW}, where we have included the sensitivity curves of various GW detectors for comparison.

\section{Results and discussion}\label{s:GWres}

\begin{figure}[ht]
    \centering
    
    \includegraphics[width=0.9\linewidth]{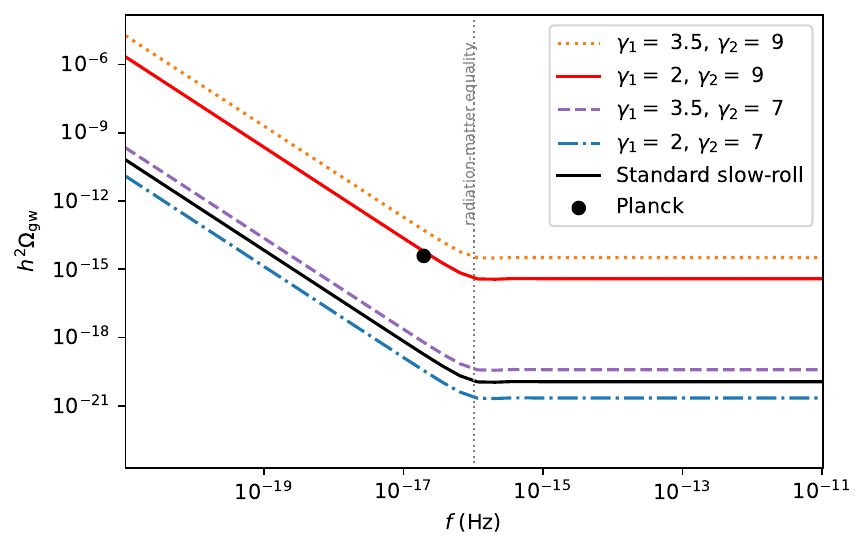}

    \caption{GW energy density at frequencies around $H_{\rm{eq}}$ for an inflation scale $T\ssend =\SI{1e15}{\GeV}$ and several values of $\gaun$ and $\ga_2$ compatible with back-reaction constraints.  Neglecting slow roll corrections, the frequency dependence is the same as the traditional inflationary tensor spectrum shown in black. The black dot indicates the upper bound from the Planck data at a pivot scale $k_* = \SI{0.002}{\mega\mathrm{pc}^{-1}}$. This is a continuation of \fig\ref{fig:omegaGW} to lower frequencies. }
    \label{fig:omGW_at_equality}
\end{figure}

We have found that gauge fields generated by kinetic and axial coupling to the inflaton generically produce a scale invariant background of gravitational waves (up to slow-roll corrections) leading to a tensor-to-scalar ratio $r(\ga_1,\ga_2)$ that strongly depends in the strength of the axial coupling parametrized by $\ga_2$, and weakly depends on the strength of the kinetic coupling parametrized by $\ga_1$. The function $h^2\Om_{\text{gw}}(\ga_1,\ga_2,f)$ therefore has nearly the same frequency dependence as GWs directly generated by inflation: see \fig\ref{fig:omGW_at_equality}, where the upper limit from the Planck experiment at the pivot scale $k_* =  \SI{0.002}{\mega\mathrm{pc}^{-1}}$ is indicated as a black dot.

However, contrary to the quantum amplification of gravitational waves, with an amplitude scaling like $(H/\planckmass)^2$, these secondary gravitational waves that are generated classically by the anisotropic stress of the quantum induced gauge fields scale as $(H/\planckmass)^4$, see \Eq\eqref{eq:final-PT}. Amplitudes of quantum fields generated during inflation are always proportional to $(H/\planckmass)^2$, and the GWs couple to the square of the field strength. Nevertheless, for sufficiently large values of $\ga_2$, the exponential factor $\propto \exp(2\pi\ga_2)$ can overcome this reduction of amplitude. One factor of $\exp(\pi\ga_2)$ is explicit in $\cosh^2(\pi\ga_2/2)$ while the other is hidden in $\De_{E,B}^2$ and becomes valid in the limit of large $\ga_2$. This is shown in \fig\ref{fig:omGW_at_equality}, where the ratio of the tensor power spectrum \eqref{eq:final-PT} generated by gauge fields to the standard inflationary one, $\PP_T^{\rm em}/\PP_T^{\rm std}$, is presented. We have taken \cite{maggiore_vol2}
\begin{equation}
    \label{eq:PT-std-inflation}
    \PP_T\supsc{std} = \frac{2}{\pi^2}\frac{H^2}{\planckmass^2} \mperiod
\end{equation}
The line $\PP_T^{\rm em}/\PP_T^{\rm std}=1$ is approximately given by
\begin{align}
\ga_2 &\simeq -1.5\log_{10}(T\ssend/\rm{GeV}) + 30 \qquad \text{(at $\gaun=2$)} \\
\ga_2 &\simeq -1.6\log_{10}(T\ssend/\rm{GeV}) + 30 \qquad \text{(at $\gaun=3.9$)}\mperiod
\end{align}

Despite the similarity between their respective spectra, we can distinguish gauge field induced gravitational waves from the usual inflationary gravitational waves in several ways. Firstly, the former are strongly polarized in the relevant regime, i.e., when $|\ga_2|>1$. In the figures we only showed $\ga_2>0$, but the resulting spectra are even in $\ga_2$ while the polarization of the generated gauge fields and gravitational waves is odd, i.e., $\PP_T^A$ changes sign with the polarization.  Secondly, gauge field induced gravitational waves are not Gaussian as their source term is the square of the Gaussian gauge field. Here, we do not explicitly calculate the bispectrum but  we expect it to be of the order $\PP_T^{3/2}$, see~\cite{Barnaby:2011vw} for the bispectrum of scalar perturbation in pure axion inflation. Finally, contrary to standard inflationary GWs, gauge field induced GWs have a small blue correction at very high frequencies, a term of order $(k\tau\ssend)^3$ coming from the contribution of the derivative $h'$ at the end of inflation through \Eq\eqref{eq:mutiple-transitions-indep-of-H0}.

\begin{figure}
    \centering
    \includegraphics[width=0.8\linewidth]{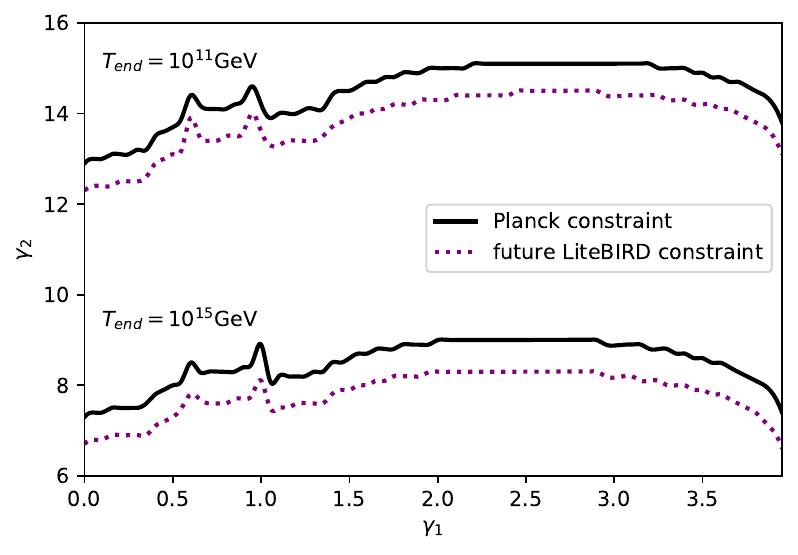}
    \caption{The regions above the lines are excluded by Planck (black) and LiteBIRD (purple, dotted), here shown for two different inflation energy scale.  We assume that the sensitivity of LiteBIRD can reach $r=10^{-3}$ \cite{LiteBIRD:2022cnt}. The visible spikes are the same that those which have been discussed in \fig\ref{fig:powerspectrum}. Note that, for clarity, the vacuum  contribution to the GW spectrum is ignored here. Parameters in the excluded region would generate an induced SGWB that already exceeds the experimental limit, and so the total SGWB including this contribution {\it and} the vacuum contribution would also be ruled out. Both contributions of the SGWB could be in principle distinguished via their non-gaussian statistics.}
    \label{fig:exclusion-plot}
\end{figure}

Finally, we discuss how observations constrain the available parameter space for $\gaun$ and $\gade$. The constraints from direct gravitational wave background searches are shown in \fig\ref{fig:omegaGW}.
However, as the initial spectrum is scale invariant, the best constraints come from very large scales tested with CMB experiments, cast in terms 
of the  tensor-to-scalar ratio bounded by the Planck experiment at the pivot scale $k_* = \SI{0.002}{\mega\mathrm{pc}^{-1}}$, namely $r_{0.002} \leqslant 0.06$ \cite{Planck:2018vyg}. Using $n_s \simeq 0.9677$ and $A_{\mathcal{R}} (\SI{0.05}{\mega\mathrm{pc}^{-1}})\simeq \SI{2.1e-9}{}$ \cite{Planck:2018vyg}, this leads to an upper bound for the tensor power spectrum at the end of inflation $\PP_{T,\text{end}}(k_*) \leqslant \SI{1.4e-10}{}$. We translate this bound into an exclusion contour in the parameter space $(\gaun,\gade)$ of our model, which is represented on \fig\ref{fig:exclusion-plot}. As a benchmark for next generation detectors, we also represent on the same figure which values of $\gaun,\gade$ could be within the reach of detection by LiteBIRD, which is planned to reach $r\simeq 10^{-3}$ \cite{LiteBIRD:2022cnt}. Note, however, that for the case of pure axion inflation other works have shown $\gade\gtrsim 6$ is excluded, as it would generate too large non-Gaussian scalar perturbations~\cite{Barnaby:2011vw}.

\section{Conclusion and outlook}\label{s:con}

Building upon previous studies, we have computed gauge fields generated by a coupling to the inflaton during slow-roll inflation. We have considered the full range of both kinetic and axial couplings within the slow-roll approximation. The kinetic coupling determines the gauge field spectral index. While the magnetic field is always blue within the considered range, the electric field becomes scale invariant for $\gaun\ra 4$.  For most parameter values, the electric field dominates, but in the vicinity of $\gaun \simeq 1$ the magnetic field can be stronger.  The axial coupling $\ga_2$ has no influence on the spectral index of the gauge fields, but it amplifies one of the two polarizations exponentially.
Its value is constrained by the condition of small back-reaction on the inflationary dynamics. In our study we have restricted $\gaun \geqslant 0$ to avoid strong coupling and $\gaun<4$ to avoid infrared divergences, but we have considered the full range $(\ga_1,\ga_2)\in \interval{[}{0}{4}{[}\times \mathbb{R}$.

After inflation the electric field is rapidly damped away by electric currents. Meanwhile, the magnetic field is damped at small scales below a time-dependent dissipation scale (see, e.g.,~\cite{Caprini:2009pr} for a study of the time dependent magnetic diffusion scale). The magnetic field spectrum remains blue,  but due to the inverse cascade in the charged cosmic plasma after inflation, it can gain sufficient large scale power to be relevant for the large scale cosmological magnetic fields discussed in the introduction, see~\cite{Caprini:2014mja} for a detailed discussion of this point.

\begin{figure}
\centering
  \includegraphics[width=0.7\linewidth]{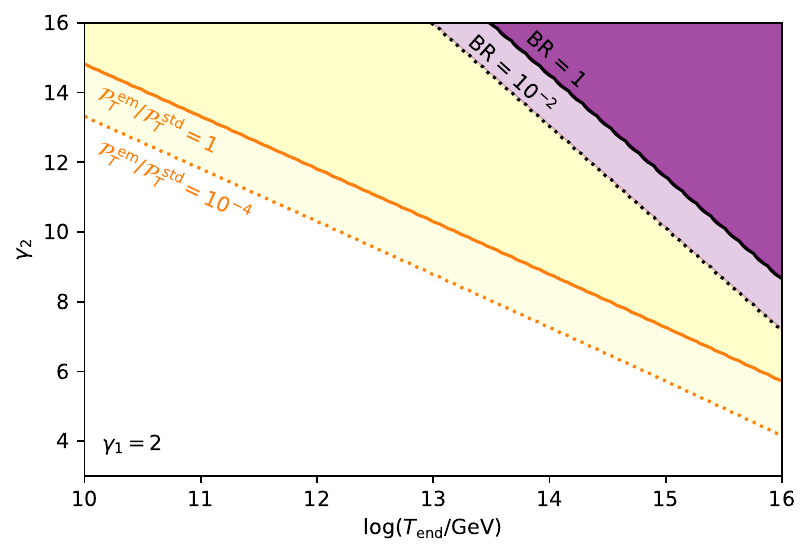}
     \includegraphics[width=0.7\linewidth]{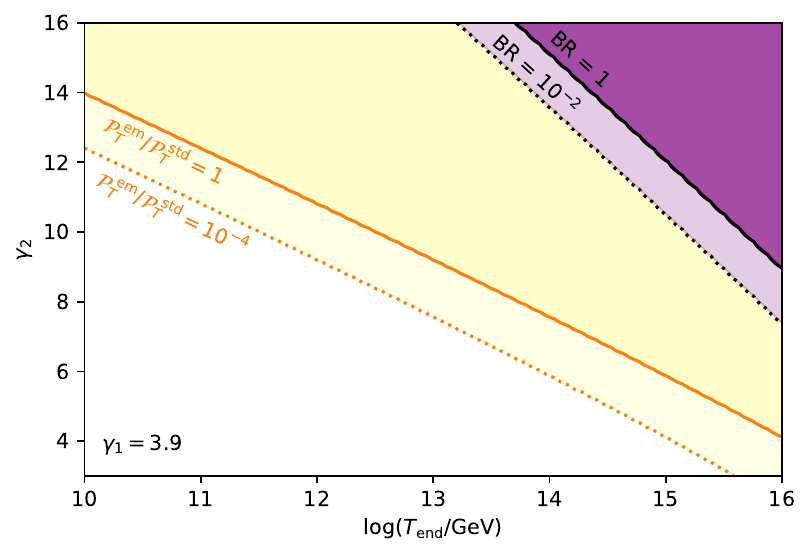}
\caption{In the yellow shaded regions of parameter space in $T\ssend$ and $\ga_2$, secondary gravitational waves generated by gauge fields are relevant but back-reaction is still unimportant. The annotation ``BR"  on the black lines shows the value of the ratio $|\partial_\phi i_2 \avg{\B\cdot\E}/(3H\dot{\phi})|$ discussed in Section\ref{sub:backreaction}. The value of $\ga_1$ chosen for the plots is indicated in the lower left corner.}
\label{f:con}
\end{figure}

Nevertheless, already during inflation, the anisotropic stress of the gauge field generates an anisotropic transverse-traceless contribution to the metric which becomes an oscillating gravitational wave at late time and is not damped subsequently. 
Depending on the couplings $(\ga_1,\ga_2)$ and on the scale of inflation set to $V^{1/4} \simeq T\ssend$, these secondary gravitational waves might actually be observable. Independent of the gauge field spectral index, the gravitational wave spectrum is scale invariant, alike the standard inflationary GW spectrum. This is a consequence of the fact that GWs are mainly produced at the horizon scale. 

This scale invariance, in fact, relies on having neglected slow-roll corrections to the expansion law of $a$. By including them we obtain a spectral index $n_T=-4\ep$ that differ from the case $n_T=-2\ep$ of standard inflationary gravitational waves \cite{maggiore_vol2}, resulting in
\be
\PP_T(k) = \left(\frac{H}{\planckmass}\right)^4\left(\frac{k}{k_*}\right)^{-4\ep}\FF(\ga_1,\ga_2)\mcomma
\ee
where $\FF(\ga_1,\ga_2)$ is exponentially growing with $\ga_2$ but depends only weakly on $\ga_1$ (as long as $\gaun$ is not too close to $4$). This difference is consistent with the general analysis of Ref. \cite{Teuscher:2025xke}. Beyond their spectral tilt, these gravitational waves can be distinguished from standard inflationary GWs in two other ways: like the gauge fields that induce them, they are strongly polarized. The polarization spectrum is given in detail in Appendix~\ref{asub:GW-comput}. Furthermore, a stochastic GW background induced by gauge fields is not Gaussian. Assuming that the gauge fields generated by vacuum amplification during inflation are Gaussian, the GWs are squares of a Gaussian field and have, e.g., a non-vanishing bispectrum. 

Moreover, for very small values of $\gade$ the induced GWs are weaker than the standard inflationary GW background, while very large values  are excluded by back-reaction. In \fig\ref{f:con} we indicate the regime of values of $H/\planckmass$ and $\ga_2$ where our secondary GW background dominates over the standard inflationary one but back-reaction is not relevant. We see that such a regime clearly exists for many inflation scales and moderately large values of the axial coupling, $\ga_2$.

Let us also mention that the limit $\gaun<4$ to avoid infrared divergence has been assumed for convenience. Without it, we would have to introduce an infrared cutoff for the electric field that would be determined by the beginning of inflation, see Ref.~\cite{Kahniashvili:2016bkp} for similar considerations. While the electric field is damped after inflation, traces of this cutoff would survive in the generated GW background. When allowing for $\ga_1 \simeq 6 $ one could then generate even scale invariant magnetic fields that would certainly be relevant for the cosmological  magnetic fields observed in the present Universe.

We often used electromagnetic terminology in the paper, it is clear that our results are valid for arbitrary $U(1)$  or other Abelian gauge fields. It is not so clear whether they can be generalized to non-Abelian fields, as has been studied in e.g. \cite{Adshead:2012kp, Maleknejad:2012fw} or more recently in \cite{Dimastrogiovanni:2023oid}.

Finally, the predicted GW background is rather on the conservative side. We have neglected the contributions after inflation of the remaining magnetic fields. We expect them, however, to be subdominant as they would be due solely to the magnetic field. Furthermore, the dominant part of the magnetic energy density that comes from small scales will be damped away by diffusion.

\section*{Acknowledgements}
The authors thank Chiara Caprini and Lucas Pinol for insightful discussions and helpful comments. M. Teuscher thanks the University of Geneva for hospitality. Part of this work was supported by the IDEX grant from the Université Grenoble-Alpes.

\appendix

\section{Equivalence between \intitle{$\phi$}-coupling and \intitle{$R$}-coupling}\label{a:phiR}

In this paper we choose to couple the $U(1)$ gauge field to the inflaton  through the action \eqref{eq:action}. Another legitimate choice would be to couple it non-minimally to gravity through a coupling to the Ricci scalar $ \propto f(R) F\munu F\upmunu$. During slow-roll inflation, these two approaches are actually equivalent. Indeed, in quasi de Sitter space the Ricci scalar is related to the Hubble parameter through $R=6(\dot{H}+2H^2) \simeq 12 H^2$. But we also have $H^2 = \flaf{1}{(3 \planckmass^2)}\rho_\phi \simeq \flaf{1}{(3 \planckmass^2)} V(\phi)$, hence $R\simeq 4V(\phi) / \planckmass^2$. Furthermore, during slow-roll the potential can be treated as a monotonic function of $\phi$, thus any function of $R$ can be translated into a function of $\phi$ and vice-versa. Hence, both formalisms are equivalent in the slow-roll approximations, where we can neglect  $\dot{H}$.

Similarly, using that to lowest order in slow-roll, the Riemann tensor can be approximated by
\begin{equation}
R_{\mu\nu\al\beta}= -H^2(g_{\mu\al}g_{\nu\beta}-g_{\nu\al}g_{\mu\beta}) \mcomma
\end{equation}
a coupling proportional to $R_{\mu\nu\al\beta}F^{\mu\nu}F^{\al\beta}$ then becomes $2H^2F^{\mu\nu}F_{\mu\nu}= (2/3)\planckmass^{-2} V(\phi)F^{\mu\nu}F_{\mu\nu}$.

\section{Notations and conventions}\label{a:conv}

Given a massless spin-$s$ field ($s=1$ or $s=2$) $\hat{X}_I(\xb,\ta)$, where $I$ collectively labels appropriate Lorentz indices (usually only spatial indices in the appropriate gauges like Coulomb gauge or transverse-traceless gauge), we write its  expansion in quantum modes as 
\begin{align}
     \label{eq:quantum-expansion-arbitrary-spin}
      \hat{X}_I(\xb,\tau) &= \kint[\kb]   e^{+i\kb\cdot\xb} \sum_{\lambda=\pm s}\hat{X}_I^{\lambda}(\kb,\ta) \\
    & = \kint[\kb] \frac{1}{\sqrt{2|\kb|}} \sum_{\lambda=\pm s}\left[\eps^\lambda_I(\kb)  X^\lambda_{\kb}(\ta) \hat{a}^\lambda_{\kb} e^{+i\kb\cdot\xb} + \text{h.c.}\right] 
    \mperiod
\end{align}
Consistency with our Fourier transform convention requires the commutation relation\\ $[\hat{a}^\lambda_{\kb},(\hat{a}^{\lambda'}_{\kb'})^\dagger] = (2\pi)^3\delta^{\lambda\lambda'}\delta^{(3)}(\kb-\kb')$. Moreover, we fix the normalization of the polarization tensors such that 
\begin{equation}
\label{eq:pol-vectors-normalization}
 \eps_I^\lambda(\kb) \eps_I^{\lambda' *}(\kb) = \delta^{\lambda\lambda'}\mcomma
\end{equation}
where a summation over the Lorentz indices $I$ is understood. The motivation behind this choice is to provide a single consistent definition of both gauge field and gravitational wave power spectra. Indeed, it implies
\begin{align}
\label{eq:def-unequaltime-correlator-general}
    \avg{\hat{X}_I^{+s}(\kb,\ta)\hat{X}_I^{+s*}(\kb',\ta') + \hat{X}_I^{-s}(\kb,\ta)\hat{X}_I^{-s*}(\kb',\ta')} &=(2\pi)^3 S_X(|\kb|,\ta,\ta')\delta^{(3)}(\kb-\kb') \\
   \avg{\hat{X}_I^{+s}(\kb,\ta)\hat{X}_I^{+s*}(\kb',\ta') - \hat{X}_I^{-s}(\kb,\ta)\hat{X}_I^{-s*}(\kb',\ta')} &=(2\pi)^3 A_X(|\kb|,\ta,\ta')\delta^{(3)}(\kb-\kb') \mcomma
\end{align}
where the (anti-)symmetric two-point functions can be obtained from the mode functions by
\begin{equation}
    \label{eq:correlation-from-mode-functions}
    S_X(k,\ta,\ta') = \frac{1}{2k}\sum_{\lambda=\pm s}X_k^\lambda(\ta)X_k^{\lambda *}(\ta') \spliteq A_X(k,\ta,\ta') = \frac{1}{2k}\sum_{\lambda=\pm s}(\lambda/s) X_k^\lambda(\ta)X_k^{\lambda *}(\ta')\mperiod
\end{equation}
The symmetric `dimensionless' power spectrum $\PP_X$ is then defined by
\begin{equation}
\avg{\hat{X}_I(\xb,\ta)\hat{X}_I(\xb,\ta)} = \int_0^\infty \frac{\dd k}{k}\PP_X(k,\ta) =  \int_0^\infty \frac{\dd k}{k}\frac{k^3}{2\pi^2}S_X(k,\ta,\ta)\mcomma
\end{equation}
and similarly
\begin{equation}
    \PP_X^A(k,\ta) = \frac{k^3}{2\pi^2}A_X(k,\ta,\ta)\mperiod
\end{equation}

\smallskip
Given an orthonormal positively oriented basis $(\hat\kb=\kb/|\kb|,\bm{u},\bm{v})$ with $\bm{v} = \hat\kb\wedge\bm{u}$, the normalization \eqref{eq:pol-vectors-normalization} is obtained for 
\begin{align}
\label{eq:def-pol-vectors-spin-1}
    \eps^\pm_j &= \frac{1}{\sqrt{2}}(u_j\pm i v_j) &\qquad \text{(spin-1)} \\
\label{eq:def-pol-vectors-spin-2}
    \eps^{\pm 2}_{ij} &= \eps^\pm_i \eps^\pm_j =\frac{1}{2}(e^+_{ij}\pm i e^{\cross}_{ij}) &\qquad \text{(spin-2)} \mperiod
\end{align}
Note that the traditional plus/cross polarization vectors 
\begin{equation}
    e^+_{ij} = u_i u_j - v_i v_j\quad,\quad e^\cross_{ij} = u_i v_j + v_i u_j
\end{equation}
do not satisfy the normalization \eqref{eq:pol-vectors-normalization} but rather $e^{\sigma = +/\cross}_{ij}e^{\sigma' =+/\cross}_{ij} = 2\delta^{\sigma\sigma'}$, hence they introduce an additional factor of 2 if their corresponding mode functions are used in \Eq\eqref{eq:correlation-from-mode-functions}. Finally, further properties of polarization tensors \eqref{eq:def-pol-vectors-spin-1} and \eqref{eq:def-pol-vectors-spin-2} include 
\begin{equation}
k_i\eps_{iJ}^\lambda(k) = 0,\quad i \epsilon_{lmq} k_m \varepsilon_{qJ}^\lambda =   (\lambda/s) |k|\varepsilon^\lambda_{lJ} \mcomma
\end{equation}
where $\epsilon_{lmq}$ is the Levi-Civita symbol, and $J$ stands for either no extra index (spin-1) or a spatial $j$ index (spin-2).

Using a stochastic formalism rather than quantum fields does not affect the normalizations in these definitions. Indeed for stochastic fields the quantum operators $\hat a_{\kb}^\la$ rather become stochastic Gaussian fields with the expectation values $\langle\hat{a}^\lambda_{\kb}\rangle =0$ and $\langle\hat{a}^\lambda_{\kb}(\hat{a}^{\lambda'}_{\kb'})^*\rangle = (2\pi)^3\delta^{\lambda\lambda'}\delta^{(3)}(\kb-\kb')$.

\section{Similarities with the Mukhanov-Sasaki equation}\label{a:mukha}

Let us recall the equation of motion \eqref{eq:eom-nice-form} for the gauge field in the special case where $\gade=0$ (we drop the polarization labels as in this case both polarizations are produced in equal amount) 

\begin{equation}
\label{eq:eom-no-gamma2}
     \cA''_k  +\left( k^2 - \frac{(\sqrt{1+i_1})''}{\sqrt{1+i_1}} \right) \cA_k =0 \mperiod
\end{equation}
This Klein-Gordon equation with time-dependent mass shares many similarities with the Mukhanov-Sasaki equation for scalar perturbations \cite{maggiore_vol2}
\begin{equation}
\label{eq:mukhanov-sasaki}
    u_k'' + \left(k^2 - \frac{z''}{z}\right)u_k = 0 \mcomma
\end{equation}
where $u_k$ is the Mukhanov-Sasaki variable, and $z\equiv a \dot{\phi}_0/H$ translates the Universe expansion, $\dot{\phi}_0$ being the small variation of the background inflaton field during slow-roll. The physical gauge field $A_k = \cA_k/\sqrt{1+i_1}$ therefore plays a role equivalent to the curvature perturbation $\mathcal{R}_k \equiv \flaf{-u_k}{z}$. However, the main difference is that at the lowest non-trivial order in slow-roll parameters, $z\propto -1/\ta$ while $\sqrt{1+i_1} = 1$. As a consequence, the mode functions of \eqref{eq:eom-no-gamma2} at lowest order, $f_k\supsc{low.}(\ta) =\frac{1}{\sqrt{2k}} e^{-ik\ta}$, do not exhibit a super-horizon regime, contrary to the De Sitter mode functions $f_k\supsc{dS}(\ta)= \frac{1}{\sqrt{2k}}\left(1-\frac{i}{k\ta}\right)$ which lead to the Harrison-Zel'dovich spectrum for $\mathcal{R}_k$. As mentioned in the main body of the paper, this is a consequence of the conformal invariance of gauge fields, that is not satisfied by the Mukhanov-Sasaki variable.

Our analysis is in fact closer to the next-to-leading-order description of scalar perturbations during slow-roll. The connection between both is made explicit in \tab\ref{tab:mukhanov}.

\begin{table}
    \centering
    \begin{tabular}{|c|c|c|}
    \hline    && \\[-0.2cm]
         & Gauge fields & Mukhanov-Sasaki variable  \\ \hline
         && \\[-0.2cm]
       Time-dependent mass  & $\dps \frac{\sqrt{1+i_1}''}{\sqrt{1+i_1}} = \frac{\nu^2-1/4}{\ta^2}$, & $\dps \frac{z''}{z} = \frac{\nu^2-1/4}{\ta^2}$, \\[0.4cm]
       & $\dps \nu - \frac{1}{2} = - \frac{\gaun}{2} \simeq \const$  & $\dps \nu - \frac{3}{2} = 3\ep-\eta\simeq \const$  \\[0.3cm] \hline
       Mode functions & Hankel function of order $\nu$ & Hankel function of order $\nu$ \\ \hline
       Spectral tilt  & $\dv{\ln \PP_A}{\ln k} = -2\left(\nu-\frac{1}{2}\right) = \gaun$ & $\dv{\ln \PP_{\mathcal{R}}}{\ln k} = -2\left(\nu-\frac{3}{2}\right) = 2\eta-6\ep$
 \\[0.2cm] \hline
   \end{tabular}
    \caption{Comparison between the results developed in Section \ref{s:electromag} and the standard analysis of slow-roll perturbations. The latter is taken from chapter 21 of \cite{maggiore_vol2} (we have introduced $\eta= \planckmass^2 (\partial^2_{\phi} V / V)$) and $\PP_A$ is the power spectrum of the gauge potential. We assume to be in the case $\gaun <1$, as $\gaun$ is here analogous to a combination of the slow-roll parameters which are small.}
    \label{tab:mukhanov}
\end{table}

\section{Details on analytical computations}
\label{a:comput}

\subsection{Electromagnetic fields}
\label{asub:EM-comput}

The functions $\delta^\pm_B$ and $\delta^\pm_E$ introduced in  Section \ref{sub:Em-spectra} are defined as
\begin{align}
    \delta^\pm_B(\gaun,\gade)&\equiv \frac{1}{\Ga\left(\frac{1}{2}+\frac{1}{2}|1-\gaun|\pm i \frac{\gade}{2}\right)}\left\{ \begin{array}{ll} 
    \Ga(|1-\gaun|) & \quad\text{if $|1-\gaun|\geqslant \delta$} \\
    -\left(2\gamma_E+\psi\left(\frac{1}{2}\pm i \frac{\gade}{2}\right) + \ln(2ik\ta)\right) & \quad\text{if $|1-\gaun|< \delta$}
    \end{array}\right.\mcomma \label{ea:delta1}\\
    \delta^\pm_E(\gaun,\gade)&\equiv \frac{1}{\Ga\left(\frac{\gaun}{2}\pm i \frac{\gade}{2}\right)}\left\{ \begin{array}{ll} 
    \pm \frac{i}{\gade}-2\gamma_E - \psi\left(\pm i \frac{\gade}{2}\right) - \ln(2ik\ta) & \quad\text{if $0\leqslant \gaun < \delta$} \\
    \Ga(\gaun) & \quad\text{if $ \gaun \geqslant \delta$} 
    \end{array}\right. \label{ea:delta2}\mperiod
\end{align}

Here, $0<\delta<1$ is a threshold value at which we switch from one approximation to the other before the divergent behavior becomes relevant. The parts involving the digamma function $\psi \equiv \Ga'/\Ga$ correspond to the finite limit reached when summing several terms of the expansion \eqref{eq:super-horizon-whittaker}, as explained in the discussion around \Eq\eqref{eq:whittaker-3rd-term}. This choice will of course leads to a discontinuity at $|1-\gaun|=\delta$ in~\eqref{ea:delta1} and at $\gaun=\delta$ in~\eqref{ea:delta2}, however this remains more accurate than keeping a divergent approximation. Based on numerical tests, in the main body of the article we set $\delta = 0.4$ unless stated otherwise. In addition, we neglect the logarithmic contribution $\ln(2ik\ta)$, so $\delta^\pm_B$ and $\delta^\pm_E$ are everywhere treated as independent of $k\ta$. This last assumption is used to define $\De_B$, $\De_E$ as
\begin{align}
\label{eq:def-DeltaB}
    \De_B(\gaun,\gade) &\equiv \frac{2^{2-|1-\gaun|}}{\left|\Ga\left(\frac{1}{2}+\frac{1}{2}|1-\gaun|+ i \frac{\gade}{2}\right)\right|^2}\times \left\{ \begin{array}{ll}
        \left|\Ga(|1-\gaun|)\right|^2 & \quad\text{if $|1-\gaun|\geqslant \delta$} \\
        \left|2\gamma_E + \psi\left(\frac{1}{2}+i\frac{\gade}{2}\right)\right|^2 & \quad\text{if $|1-\gaun|< \delta$}
    \end{array}\right. \mcomma \\
    \label{eq:def-DeltaE}
     \De_E(\gaun,\gade) &\equiv \frac{2^{3-\gaun}}{\left|\Ga\left(\frac{\gaun}{2}+ i \frac{\gade}{2}\right)\right|^2}\times \left\{ \begin{array}{ll}
        \left|2\gamma_E + \psi\left(i\frac{\gade}{2}\right)-\frac{i}{\gade}\right|^2 & \quad\text{if $0\leqslant \gaun < \delta$} \\
         \left|\Ga(\gaun)\right|^2 & \quad\text{if $\gaun\geqslant \delta$} 
    \end{array}\right. \mperiod
\end{align}
We have used $\Ga(z^*)=\Ga(z)^*$ and $\psi(z^*)=\psi(z)^*$ to remove some dependences on the polarization, allowing to write \eqref{eq:S-EB} {\it et seq.} in a concise factorized form. Note that for $ \left|\beta\right| \gg \left|\al\right| $, 
$\left|\Ga(\al+i\beta)\right| \simeq \exp(-\pi|\beta|/2) |\beta|^\alpha\sqrt{2\pi|\beta|}$. For large $\ga_2$ both, $\De_E$ and $\De_B$ therefore scale like $\exp(\pi\ga_2/2)$ up to power law corrections.

To simplify the back-reaction calculation we first recall that our definitions in Appendix~\ref{a:conv} and section~\ref{s:electromag} lead to the following expectations values for the second moments of the gauge fields:
\begin{align}
\avg{\bm{\B}^2} &= \int_0^{k_h} \frac{\dd k}{k}\PP_\B(k) =  (1+i_1) \sum_{\lambda=\pm} \int_0^{k_h} \frac{\dd k}{k} \frac{k^5}{2\pi^2 a^4}\pref{2k}\left|\frac{\cA_k^\lambda}{\sqrt{1+i_1}}\right|^2 \\
\avg{\bm{\E}^2} &= \int_0^{k_h} \frac{\dd k}{k}\PP_\E(k) = (1+i_1) \sum_{\lambda=\pm} \int_0^{k_h} \frac{\dd k}{k} \frac{k^3}{2\pi^2 a^4}\pref{2k}\left|\dv{}{\ta}\left(\frac{\cA_k^\lambda}{\sqrt{1+i_1}}\right)\right|^2 \\
 \avg{\bm{\B}\cdot\bm{\E}} &= \sum_{\lambda=\pm} \int_0^{k_h} \frac{\dd k}{k} \frac{k^3}{2\pi^2}\pref{2k}\left(\B_k^{\lambda} \E_k^{\lambda*}+\B_k^{\lambda*}\E_k^{\lambda}\right)\mperiod 
\end{align}

\begin{figure}
    \centering
    \includegraphics[width=0.49\linewidth]{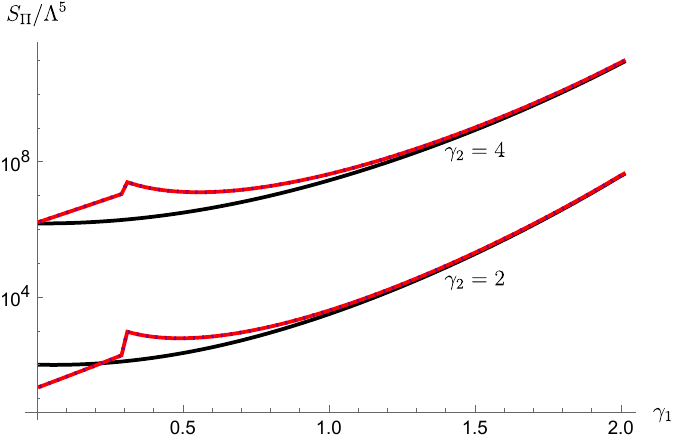}  \includegraphics[width=0.49\linewidth]{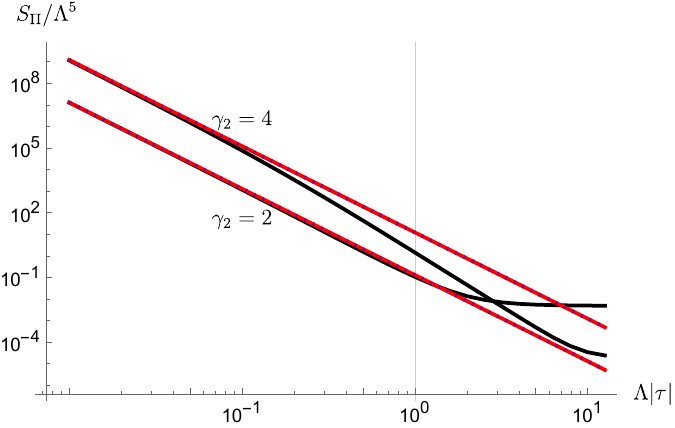}
    \caption{The symmetric spectrum $S_\Pi$ for various values of $\gaun,\gade$. Solid black lines correspond to the numerical integral of the exact Whittaker functions, while solid red line correspond to our analytical expression \eqref{eq:final-S-Pi}. To obtain the dotted purple lines (that is perfectly covered by the red lines) we expanded the Whittaker function but performed the integral \eqref{eq:angular-integral} numerically, which shows that the formulae \eqref{eq:angular-integral-symS}--\eqref{eq:angular-integral-symA} are correct. Furthermore, one observes an strong dependence on $\gade$ that is due to the exponential enhancement of one polarization when $\gade$ is non-zero. In the left (resp. right) panel we have set $|\tau|=10^{-2}/\Lambda$ (resp. $\gaun = 2$), and  $k=0$, $\Delta_B=0$ (see the text). The vertical line marks the value $\Lambda|\ta|=\gatr$.}
    \label{fig:anisotropy-spectra}
\end{figure}

\bigskip
In \fig\ref{fig:anisotropy-spectra} we compare \Eqs \eqref{eq:final-S-Pi} and \eqref{eq:final-A-Pi} with the numerical integration of the exact Whittaker functions. As it is dominant in most phenomenologically relevant situations we focus on the contribution of the electric field only by taking $\De_B= 0$ in the aforementioned equations. The numerical convergence of the 3D-integral \eqref{eq:aniso-correl-equaltime} when inserting the exact Whittaker functions turns out to be a difficult problem; however it is possible to circumvent this issue. Indeed, the integral of a blue spectrum up to a UV-cutoff $\Lambda$  results in a white noise spectrum, if the numerical and analytical integrations match at a given $k\ll\Lambda$, they should hence match at any $k$ below this cutoff. We therefore use the limit $k\to 0$ of \eqref{eq:final-S-Pi} for which the angular part of the integral simplifies considerably,\footnote{When $k=0$ the antisymmetric spectrum $A_\Pi$ vanishes by symmetry arguments, so this trick fails. For this reason, here we focus on the symmetric spectrum $S_\Pi$ only.} making the numerical integration of the Whittaker function possible. The result is illustrated in \fig\ref{fig:anisotropy-spectra}, and we conclude that our analytical expression based on expanding the Whittaker function is very close to the numerical estimate as long as $\gaun$ is not too small.

\subsection{Source of the gravitational wave spectrum}
\label{asub:GW-comput}

\begin{figure}
    \centering
    \includegraphics[width=0.8\linewidth]{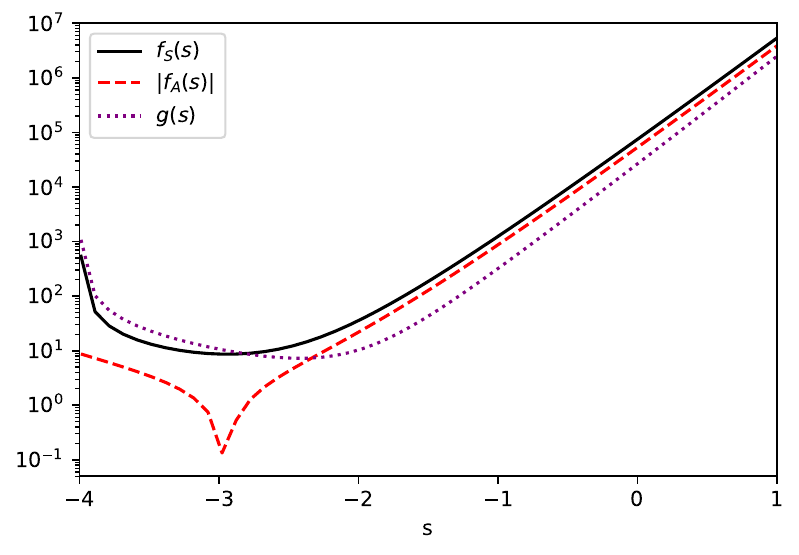}
    \caption{Values of the functions $f_S$, $f_A$ and $g$ in the case $\La/k = 10$. As $0\leqslant \gaun < 4$ and $s$ is either $-\gaun$, $1-|1-\gaun|$ or their average, it runs from $-4$ to $1$. The growth observed for $s\geqslant 5/2$ corresponds to the regime where $f_S, f_A, g$ scale as $\La/k$  to some positive power. The absolute value of $f_A$ is here shown as it changes sign around $s\simeq -3$.}
    \label{fig:fSfAg}
\end{figure}

To obtain the power spectrum of $h_{ij}$ and $h'_{ij}$ we first determine the spectrum of $\Pi_{ij}$. The starting point are equations \eqref{eq:angular-integral-symS}--\eqref{eq:angular-integral-antisym}. We define 
\begin{align}
  f_S(\La,k,s)&=\int_0^\La \frac{\dd p}{p} p^{4+s}\int_{-1}^1\dd\mu (k^2+p^2 -2k p\mu)^{\frac{1+s}{2}}(1+\mu^2)(1+\bt^2) \\
 f_A(\La,k,s)&=\int_0^\La \frac{\dd p}{p} p^{4+s}\int_{-1}^1\dd\mu (k^2+p^2 -2k p\mu)^{\frac{1+s}{2}}4\mu\bt \\
 g(\La,k,s)&=\int_0^\La \frac{\dd p}{p} p^{4+s}\int_{-1}^1\dd\mu (k^2+p^2 -2k p\mu)^{\frac{1+s}{2}}4(1+\mu^2)\bt \mcomma
\end{align}
where  $s\in\interval{]}{-4}{1}{]}$ is a function of $\ga_1$, and we recall that $\mu = \hat{\bm{k}}\cdot\hat{\bm{p}}$, $\bt = \hat{\bm{k}}\cdot\widehat{\bm{k}-\bm{p}}= \frac{k-p\mu}{\sqrt{k^2+p^2-2kp\mu}}$. We next split $\int_0^\La = \int_0^k + \int_k^\La$, which allows to expand the integrand in  a power series of $p/k$ (resp. $k/p$) in the first (resp. second) integral. After gathering all terms into a unique power series on each side, we obtain for any $f\in\{f_S,f_A,g\}$
\begin{equation}
\label{eq:concise-expr-fSfAg}
    f(\La,k,s) = k^{5+2s}\left[\xi^f(s) + \sum_{n=0}^\infty\frac{1}{5+2s-n}\left(\left(\frac{\La}{k}\right)^{5+2s-n}-1\right) c_n^f(s)\right] \mperiod
\end{equation}
Note that for large values of $\ga_2$ where $\La =1/(-\tau)$, the functions $f$ are independent of $\ga_2$.
To define the coefficients $\xi^f$ and $c_n^f$, we first introduce terms related to the power series of $z\mapsto(1+z)^{\alpha}$, that is,
\begin{equation}
   \forall\ell\in\mathbb{\mathbb{N}},\quad a_\ell^{(\alpha)}(\mu) =\sum_{j= \lfloor(\ell+1)/2\rfloor}^\ell \binom{\alpha/2}{j}\binom{j}{\ell-j}(-2\mu)^{2j-\ell} \mcomma
\end{equation}
where $\binom{x}{j} \equiv \frac{\Ga(x+1)}{\Ga(j+1)\Ga(x-j+1)}$. With the convention that $a_\ell^{(\alpha)} = 0$ if $\ell <0$, we can express the coefficients defined in \Eq\eqref{eq:concise-expr-fSfAg} as (we drop the $\mu$-dependence of $a_\ell^{(s)}$ for the sake of conciseness)
\begin{align}
     \xi^{f_S}(s) &= \sum_{n=0}^\infty \frac{1}{4+s+n}\int_{-1}^1\dd \mu\, (1+\mu^2)\left(a_n^{(s+1)}+a_n^{(s-1)} -2\mu a_{n-1}^{(s-1)} + \mu^2 a_{n-2}^{(s-1)}\right) \\
      \xi^{f_A}(s) &= \sum_{n=0}^\infty \frac{1}{4+s+n}\int_{-1}^1\dd \mu\, 4\mu\left(a_n^{(s)}-\mu a_{n-1}^{(s)} \right) \\
       \xi^{g}(s) &= \sum_{n=0}^\infty \frac{1}{4+s+n}\int_{-1}^1\dd \mu\, 4(1+\mu^2)\left(a_n^{(s)}-\mu a_{n-1}^{(s)} \right)  \\
       c^{f_S}_n(s) &= \int_{-1}^1\dd \mu\, (1+\mu^2)\left(a_n^{(s+1)}+\mu^2 a_n^{(s-1)} -2\mu a_{n-1}^{(s-1)} + a_{n-2}^{(s-1)}\right) \\
      c^{f_A}_n(s) &= \int_{-1}^1\dd \mu\, 4\mu\left(-\mu a_n^{(s)} + a_{n-1}^{(s)} \right)  \\
       c^{g}_n(s) &= \int_{-1}^1\dd \mu\, 4(1+\mu^2)\left(-\mu a_n^{(s)} + a_{n-1}^{(s)} \right)  \mperiod
\end{align}
We emphasize that, since $z\mapsto (1+z)^\alpha$ has a convergence radius of (at least) $1$, \Eq\eqref{eq:concise-expr-fSfAg} provides the {\it exact} expressions for $f_S$, $f_A$ and $g$. We further note that since $a_0^{(s)}=1 $, we obtain $c^g_0(s) = 0$, so for $f=g$ the series in the latter equation can be indexed starting at $n=1$. We have verified that these series usually converge rapidly, making these equations useful in practice. Typical values for $f_S$, $f_A$ and $g$ are depicted on \fig\ref{fig:fSfAg}. 

We can now compute the GW power spectrum  using \eqref{eq:PT-from-double-int} {\it et seq.}, inserting \Eq\eqref{eq:concise-expr-fSfAg} in \eqref{eq:final-S-Pi}--\eqref{eq:final-A-Pi}. Writing the UV cutoff as
\begin{equation}
    \La(\ta',\ta'') = \frac{k\gatr}{\max(-y,-z)}\mcomma
\end{equation}
with $y=k\ta',z=k\ta''$, we find that (all integrals run from $-\gatr$, horizon crossing, to $x=k\ta$ deep in the super-horizon regime, $x\ra 0^-$)
\begin{align}
\label{eq:formula-I-sigtau}
    &\int\!\!\! \int x^2 G(x,y)G(x,z) y^{1+s}z^{1+r}\dd y\dd z \simeq \frac{\gatr^{8+s+r}}{9(4+s)(4+r)} \simeq  \int\!\!\!\int G_1(x,y)G_1(x,z)y^{1+s}z^{1+r}\dd y\dd z \mcomma\\
    \label{eq:formula-J-sig}
    &\int\!\!\!\int x^2 G(x,y)G(x,z)(yz)^{1+s}\left[\frac{\gatr^{5+2s-n}}{\max(-y,-z)^{5+2s-n}} -1\right]\dd y\dd z \simeq \frac{(5+2s-n)}{9(4+s)^2(3+n)}\gatr^{8+2s} \mcomma\\
    \label{eq:formula-J-tau}
    &\int\!\!\!\int G_1(x,y)G_1(x,z)(yz)^{1+s}\left[\frac{\gatr^{5+2s-n}}{\max(-y,-z)^{5+2s-n}} -1\right]\dd y\dd z \simeq \nonumber \\
    &  \hspace{6.15cm}
       \frac{\gatr^{5+2s}}{|k\ta|} \frac{(5+2s-n)(9\cdot \mathbbm{1}_{n=0}+|k\ta|\gatr^3\cdot \mathbbm{1}_{n> 0})}{9(4+s)^2(3+n)} \mperiod
\end{align}
For these results we have approximated the Green functions in the limit $x,y\ll 1$, however we have assessed that if one uses the full expression \eqref{eq:Green-from-sin-and-cos} instead of relying on this assumption, the numerical value of the integral is very close to these analytical estimates as long as $\gatr \lesssim 1$. The integrand is always dominated by the $-\gatr$ bound, rendering the results independent of both $k$ and $\ta$. The only  exception is \Eq\eqref{eq:formula-J-tau} for $n=0$, in which case the integral is $\propto 1/|k\ta|$. 
The cross-term involving a contribution from $B$ and from $E$ can be computed in a similar fashion noting that
\bea
    \int\!\!\!\int \sig(y,z)\frac{(-y)^s(-z)^r+(-y)^r(-z)^s}{\max(-y,-z)^{5+s+r-n}} &=&  \nonumber \\
    &&  \hspace{-5cm}\int\!\!\!\int \sig(y,z)\frac{(yz)^s}{\max(-y,-z)^{5+2s-n}} + \int\!\!\!\int \sig(y,z)\frac{(yz)^r}{\max(-y,-z)^{5+2r-n}} \mcomma
\eea
for a function $\sigma$ symmetric in its two arguments. 

The last step is now to combine the time and angular integration. The key observation is that the factors $5+2s-n$ present in both \eqref{eq:concise-expr-fSfAg} and \eqref{eq:formula-J-sig}--\eqref{eq:formula-J-tau} cancel, so that the resulting power spectrum is divergence-free, as argued at the end of Section \ref{subsub:aniso-cutoff}.  Defining the expressions
\begin{equation}
    C(f,s) = \xi^{f}(s) + \sum_{n=0}^\infty \frac{c_n^{f}(s)}{3+n} \quad , \quad  C_1(f,s) = C(f,s) - \frac{1}{3}c_0^f(s) \mcomma
\end{equation}
we finally obtain the following GW power spectra
\begin{align}
\nonumber
 \PP_T
     =& \frac{H^4}{\planckmass^4} \FF(\gaun,\gade) \\
     =& \frac{H^4}{\planckmass^4} \frac{\gatr^8}{2\pi^2(4\pi)^2}\left[\vphantom{\frac{\pi\gade}{3}}\frac{\De_B^2 \gatr^{2 m_B}}{9(4+m_B)^2}\left(\cosh^2\left(\frac{\pi\gade}{2}\right) C(f_S,m_B) \right. \right. \nonumber \\
    & \left. \hspace{4.5cm}+ \sinh^2\left(\frac{\pi\gade}{2}\right)C(f_A,m_B)\right) 
    \nonumber  \\ 
 +& \frac{\De_E^2 \gatr^{2 m_E}}{9(4+m_E)^2}\left(\cosh^2\left(\frac{\pi\gade}{2}\right) C(f_S,m_E) + \sinh^2\left(\frac{\pi\gade}{2}\right)C(f_A,m_E)\right)  \nonumber \\ 
  +&  \frac{2\De_B\De_E \gatr^{m_B+m_E}}{9(4+m_B)(4+m_E)}\left(\cosh^2\left(\frac{\pi\gade}{2}\right) C\!\left(f_S,\frac{m_B+m_E}{2}\right) \right. \nonumber \\
    &  \hspace{3.2cm}+\left.\left. \sinh^2\left(\frac{\pi\gade}{2}\right)C\!\left(f_A,\frac{m_B+m_E}{2}\right)\right) \right]
       \label{eq:full-PT}
\\
\nonumber
\PP^A_T
     =&  \frac{H^4}{\planckmass^4} \FF_A(\gaun,\gade)\\
     =& \frac{H^4}{\planckmass^4} \frac{\gatr^8}{2\pi^2(4\pi)^2}\frac{\sinh\pi\gade}{2} \left[ 
      \frac{\De_B^2 \gatr^{2 m_B}}{9(4+m_B)^2} C(g,m_B)
      \frac{\De_E^2 \gatr^{2 m_E}}{9(4+m_E)^2}C(g,m_E)   \right.  \nonumber \\
  &\hspace*{4cm} +\left. 
  \frac{2\De_B\De_E \gatr^{m_B+m_E}}{9(4+m_B)(4+m_E)} C\!\left(g,\frac{m_B+m_E}{2}\right)\right]
       \label{eq:full-PAT}  
\end{align}
\begin{align}
\nonumber
     \Om\ssgw
     =&  \frac{H^4}{\planckmass^4} (-k\ta)^3\FF'(\gaun,\gade,k\ta)\\
     =& \frac{H^4}{\planckmass^4}(-k\ta)^3 \frac{\gatr^5}{24\pi^2(4\pi)^2}\left[\vphantom{\frac{\pi\gade}{3}} \frac{\De_B^2 \gatr^{2 m_B}}{9(4+m_B)^2}\left(\cosh^2\left(\frac{\pi\gade}{2}\right) (3+(-k\ta)\gatr^3 C_1(f_S,m_B))  \right. \right. \nonumber\\ 
  \nonumber  & \hspace{5.8cm}+ \left. \sinh^2\left(\frac{\pi\gade}{2}\right)(3+(-k\ta)\gatr^3 C_1(f_A,m_B))\right) 
   \\ 
  \nonumber +&  \frac{\De_E^2 \gatr^{2 m_E}}{9(4+m_E)^2}\left(\cosh^2\left(\frac{\pi\gade}{2}\right) (3+(-k\ta)\gatr^3 C_1(f_S,m_E)) \right. \\ 
  \nonumber  & \hspace{1.8cm}+ \left. \sinh^2\left(\frac{\pi\gade}{2}\right)(3+(-k\ta)\gatr^3 C_1(f_A,m_E))\right)  \\ 
  \nonumber
  +&  \frac{2\De_B\De_E \gatr^{m_B+m_E}}{9(4+m_B)(4+m_E)}\left(\cosh^2\left(\frac{\pi\gade}{2}\right) \left(3+(-k\ta)\gatr^3 C_1\!\left(f_S,\frac{m_B+m_E}{2}\right)\right)\right. \\
  & \hspace*{3.5cm}\left.\left.+ \sinh^2\left(\frac{\pi\gade}{2}\right)\left(3+(-k\ta)\gatr^3 C_1\!\left(f_A,\frac{m_B+m_E}{2}\right)\right)\right) \right]
       \label{eq:full-OmGW}
\\
\nonumber
     \Om\ssgw^A
     =&  \frac{H^4}{\planckmass^4}(-k\ta)^4 \FF'_A(\gaun,\gade) = \frac{1}{12}(-k\ta)^4\PP_T^A
       \label{eq:full-OmAGW} \mperiod
\end{align}

We again emphasize that $\Om\ssgw,\Om^A\ssgw$ do not represent actual energy densities, but are a way to express the typical value of the derivative $h_k^{\pm\prime}$ during inflation, which affects the initial condition for the GWs propagating in subsequent cosmological eras, see Section \ref{sub:other-eras}.

The terms inside the square brackets are functions of $\ga_1$ and $\ga_2$ only, over which we have full control with our analytical approximation. They depend exponentially on $\ga_2$ and rather weakly on $\ga_1$. Note that apart from the terms $\cosh^2(\pi\ga_2/2)$ and 
$\sinh^2(\pi\ga_2/2)$ which grow like $\exp(\pi\ga_2)$ for large 
$\ga_2$, also $\De_B$ and $\De_E$ grow like $\exp(\pi\ga_2/2)$ so that for large values of  $\ga_2$ the spectra grow like $\exp(2\pi\ga_2)$, in agreement with previous studies~\cite{Caprini:2014mja}.

\bibliographystyle{JHEP}
\bibliography{references}

\end{document}